\documentclass[10pt,preprint]{emulateapj}  % For more compact output
\usepackage{subfigure,verbatim}

\newcommand{\msc}[1]{\nobreak{\mbox{\scriptsize{\textsc{#1}}}}} % Ditto, but for small caps
\newcommand{\cross}{\times}
\newcommand{\bmath}[1]{\mbox{\boldmath{$#1$}}}

\def\deg{^\circ}
\def\comp{\,c/\omega_{\rm pi}}
\def\ompt{\omega_{\rm pi}t}
\def\omp{\,\omega_{\rm pi}^{-1}}
\def\thetacrit{\theta_{\rm crit}}

\newcommand{\gbi}[1]{\gamma_i\beta_{{\rm #1}i}}

\newcommand{\tit}[1]{\textit{#1}}

\newcommand{\fig}[1]{Fig.~\ref{fig:#1}}
\newcommand{\fign}[1]{\ref{fig:#1}}

\newcommand{\pow}[1]{10^{-{#1}}}
\newcommand{\ditto}[1]{{_{\rm{#1}}}} % Ditto, but for small caps
\newcommand{\crs}{_{\nobreak{\mbox{\scriptsize{\textsc{cr}}}}}} % Ditto, but for small caps
\begin{document}
\title{Particle Acceleration in Relativistic Magnetized Collisionless Electron-Ion Shocks}
\author{Lorenzo Sironi and Anatoly Spitkovsky}
\affil{Department of Astrophysical Sciences, Princeton University, Princeton, NJ 08544-1001, USA}
\email{lsironi@astro.princeton.edu;\\ anatoly@astro.princeton.edu}

\begin{abstract}
We investigate shock structure and particle acceleration in relativistic magnetized collisionless electron-ion shocks by means of 2.5D particle-in-cell simulations with ion-to-electron mass ratios ($m_i/m_e$) ranging from 16 to 1000. We explore a range of inclination angles between the pre-shock magnetic field and the shock normal. In ``subluminal'' shocks, where relativistic particles can escape ahead of the shock along the magnetic field lines, ions are efficiently accelerated via a Fermi-like mechanism. The downstream ion spectrum consists of a relativistic Maxwellian and a high-energy power-law tail, which contains $\sim5\%$ of ions and $\sim30\%$ of ion energy. Its slope is $-2.1\pm0.1$. The scattering is provided by short-wavelength non-resonant modes produced by Bell's instability (Bell 2004), whose growth is seeded by the current of shock-accelerated ions that propagate ahead of the shock. Upstream electrons enter the shock with lower energy than ions (albeit by only a factor of $\sim5\ll m_i/m_e$), so they are more strongly tied to the field. As a result, only $\sim1\%$ of the incoming electrons are Fermi-accelerated at the shock before being advected downstream, where they populate a steep power-law tail (with slope $-3.5\pm0.1$). For ``superluminal'' shocks, where relativistic particles cannot outrun the shock along the field, the self-generated turbulence is not strong enough to permit efficient Fermi acceleration, and the ion and electron downstream spectra are consistent with thermal distributions. The incoming electrons are heated up to equipartition with ions, due to strong electromagnetic waves emitted by the shock into the upstream. Thus, efficient electron heating ($\gtrsim15\%$ of the upstream ion energy) is the universal property of relativistic electron-ion shocks, but significant nonthermal acceleration of electrons ($\gtrsim2\%$ by number, $\gtrsim10\%$ by energy, with slope flatter than $-2.5$) is hard to achieve in magnetized flows and requires weakly magnetized shocks (magnetization $\sigma\lesssim\pow{3}$), where magnetic fields self-generated via the Weibel instability are stronger than the background field. These findings place important constraints on the models of AGN jets and Gamma Ray Bursts that invoke particle acceleration in relativistic magnetized electron-ion shocks.
\end{abstract}


\section{Introduction}\label{sec:intro}
Nonthermal emission from Pulsar Wind Nebulae (PWNe), jets from Active Galactic Nuclei (AGNs), Gamma-Ray Bursts (GRBs) and supernova remnants (SNRs) is usually modeled as synchrotron or inverse Compton radiation from a power-law population of electrons accelerated at collisionless shocks. The slope of the power-law tail and the acceleration efficiency, i.e., the fraction of particles and energy stored in the tail, are usually parameterized \textit{ad hoc} to fit the observations, due to the lack of a fully self-consistent theory of particle acceleration in shocks. 
 
When encountering a collisionless shock, charged particles may be accelerated by means of two basic mechanisms. In first-order Fermi acceleration (or Diffusive Shock Acceleration, DSA), particles stochastically diffuse back and forth across the shock front and gain energy by scattering from magnetic turbulence embedded in the converging flows \citep[e.g.,][]{blandford_ostriker_78, bell_78,drury_83,blandford_eichler_87}. Alternatively, in magnetized flows, particles may also gain energy directly from the background motional  electric field $\mathbf{E}=-\mbox{\boldmath{$\beta$}}\times\mathbf{B}$ while they gyrate around the shock. The latter process is given different names depending on the barrier that reflects the particles back upstream, thereby allowing for multiple energizations. If the reflecting barrier is magnetic, due to the shock compression of the upstream field, the acceleration mechanism is called Shock Drift Acceleration  \citep[SDA; e.g.,][]{chen_75,webb_83,begelman_kirk_90}. If the barrier is electrostatic, caused by electron-ion charge separation at the shock, the process is named Shock Surfing Acceleration  \citep[SSA; e.g.,][]{lee_96, hoshino_shimada_02,shapiro_03}.
The SSA mechanism can only operate in electron-ion shocks (and not in electron-positron shocks), since no electrostatic barrier appears if the incoming species have the same rigidity.  In general, not only the elemental composition, but also the magnetization and bulk Lorentz factor of the upstream flow may be important in determining the mechanism responsible for particle energization and the resulting acceleration efficiency. If the upstream medium is magnetized, an additional parameter is the obliquity angle that the upstream magnetic field makes with the shock direction of propagation. 

In the absence of turbulence, cold particles are constrained to slide along the ordered magnetic field lines, which are advected downstream from the shock.  For high magnetic inclinations, in order to return upstream the particles should be moving along the field faster than the speed of light \citep{begelman_kirk_90}. If efficient acceleration requires repeated crossings of the shock, such ``superluminal'' geometries should be very inefficient for particle acceleration \citep{kirk_heavens_89, ballard_heavens_91}. This conclusion can be questioned in the following cases: (\tit{i}) a fraction of the incoming particles is pre-heated such that their gyro-radius is larger than the thickness of the shock, and they can experience multiple shock crossings before being advected downstream;  (\tit{ii}) there are appreciable fluctuations of the magnetic field at the shock, that may create local ``subluminal'' configurations where particle acceleration can occur.
In both cases, the final outcome is difficult to predict. In support of the pre-heating scenario, a candidate mechanism has been proposed for electron heating in magnetized oblique shocks. The synchrotron maser instability at relativistic shock fronts \citep{langdon_88} generates a coherent train of intense electromagnetic ``precursor'' waves propagating upstream, which push on electrons and make them lag behind ions. The difference in bulk velocities between electrons and ions generates longitudinal electrostatic oscillations \citep{lyubarsky_06}, that eventually dissipate by heating the upstream electrons \citep{hoshino_08}. However, it is not clear if this will result in electron acceleration, besides heating. Moreover, if the incoming electrons are too hot, the synchrotron maser instability may be suppressed, thereby self-limiting the whole process.

Alternatively, as mentioned above in (\tit{ii}), particle acceleration in superluminal shocks may be permitted if a sufficiently strong turbulence reorients the field at the shock, opening subluminal channels where particles can return upstream. If the pre-shock medium is not turbulent by itself, such magnetic field fluctuations need to be generated by shock-accelerated particles that escape upstream across field lines. However, the existence of these particles is not obvious \tit{a priori}, since  in turn it requires  sufficient magnetic turbulence to mediate their acceleration. In any case, the highly nonlinear problem of wave generation and particle heating and acceleration needs to be addressed with a self-consistent approach. 

Fully kinetic particle-in-cell (PIC) simulations provide a powerful tool for exploration of the structure of collisionless shocks from first principles, thus determining self-consistently the interplay between shock-generated waves and accelerated particles. As opposed to semi-analytic kinetic theory methods \citep[e.g.,][]{kirk_heavens_89, ballard_heavens_91,achterberg_01,keshet_waxman_05} or Monte Carlo test-particle simulations \citep[e.g.,][]{ostrowski_bednarz_98, niemiec_ostrowski_04, ellison_double_04, baring_10}, PIC simulations can tackle the problem of particle acceleration in shocks without the need for simplifying assumptions about the nature of the magnetic turbulence or the details of wave-particle interactions. 

Multi-dimensional PIC simulations of relativistic \tit{unmagnetized} shocks have been presented by \citet{spitkovsky_05} for electron-positron flows \citep[see also][]{chang_08,keshet_09,haugbolle_10}, and by \citet{spitkovsky_08} for electron-ion flows. \citet{spitkovsky_08b} and \citet{martins_09} have shown, respectively for electron-positron and electron-ion plasmas, that unmagnetized shocks naturally produce accelerated particles as part of the shock evolution. A comprehensive study of particle acceleration in relativistic \tit{magnetized}  \tit{electron-positron} shocks has been performed by \citet[hereafter SS09]{sironi_spitkovsky_09} by means of multi-dimensional PIC simulations. In agreement with earlier one-dimensional (1D) results \citep{langdon_88,gallant_92}, they found that particle acceleration in superluminal pair shocks is extremely inefficient, meaning that self-generated turbulence is not strong enough to allow for significant particle diffusion across field lines.

In this work, we extend the analysis of SS09 by investigating via 2.5D PIC simulations the properties of relativistic \tit{magnetized electron-ion} shocks. In particular, we explore how the level of electron heating (namely, the fraction of pre-shock bulk kinetic energy transferred to post-shock electrons) and the efficiency of particle acceleration depend on the upstream bulk Lorentz factor, magnetization, and magnetic obliquity. We find that electrons are heated up to a few tens of percent of the upstream ion energy, regardless of the upstream conditions, and they can even reach equipartition with ions in high-obliquity magnetized shocks. With regards to particle acceleration, efficient nonthermal energization of ions (and electrons, to a lesser degree) via a Fermi-like process is observed for \tit{subluminal} configurations. The pressure of shock-accelerated ions propagating ahead of the shock can substantially perturb the incoming flow and alter the shock structure. In contrast, the downstream spectrum of both ions and electrons in \tit{superluminal} shocks does not show any signature of acceleration to nonthermal energies, in agreement with the results of SS09 for electron-positron shocks. 

This work is organized as follows. In \S\ref{sec:setup} we discuss the setup of our simulations and the magnetic field geometry. In \S\ref{sec:fluid} the shock structure and internal physics is investigated for one representative subluminal and one superluminal obliquity; for the same angles, in \S\ref{sec:accel} we explore the mechanisms responsible for particle heating and acceleration. The main results of our work, concerning the efficiency of electron heating and particle acceleration as a function of bulk Lorentz factor, magnetization and magnetic obliquity, are presented in \S\ref{sec:survey}. We summarize our findings in \S\ref{sec:disc} and comment on the application of our results to astrophysical scenarios.

%%%%%%%%%%%%%%%%%%%%%%%%%%%%%%%%%%%%%%%%%
\section{Simulation Setup}\label{sec:setup}
We use the 3D electromagnetic PIC code TRISTAN-MP \citep{spitkovsky_05}, which is a parallel version of the publicly available code TRISTAN \citep{buneman_93} that was optimized for studying collisionless shocks. Our simulation setup parallels SS09 very closely, which we repeat here for completeness.

\begin{figure}[tbp]
\begin{center}
\includegraphics[width=0.5\textwidth]{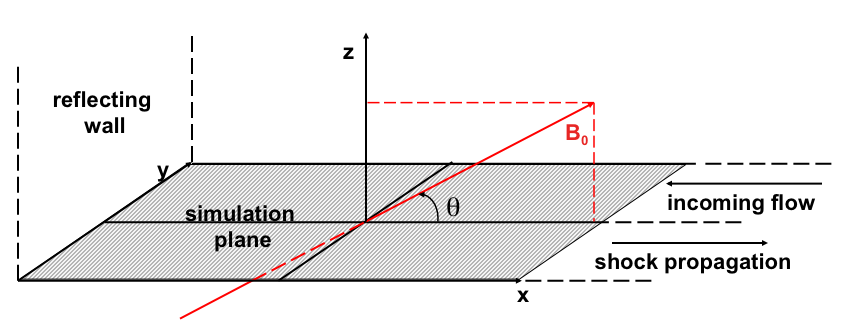}
\caption{Simulation geometry. The magnetic field (red arrow) can be either out of the simulation plane (as sketched here) or in the plane.}
\label{fig:simplane}
\end{center}
\end{figure}

The shock is set up by reflecting a cold electron-ion ``upstream''  flow from a conducting wall located at $x = 0$ (\fig{simplane}). The interaction between the incoming beam (that propagates along $-\bmath{\hat{x}}$) and the reflected beam triggers the formation of a shock, which moves away from the wall along $+\bmath{\hat{x}}$. This setup is equivalent to the head-on collision of two identical plasma shells, which would form a forward and reverse shock and a contact discontinuity. Here, we follow only one of these shocks, and replace the contact discontinuity with the conducting wall. 
The simulation is performed in the ``wall'' frame, where the ``downstream'' plasma behind the shock has zero $x$-velocity. Since the downstream plasma is at rest, no extra boosts are needed to study the spectra.

We perform simulations in both 2D and 3D computational domains, and we find that most of the shock physics is well captured by 2D simulations. Therefore, to follow the shock evolution for longer times with fixed computational resources, we mainly utilize 2D runs. All three components of particle velocities and electromagnetic fields are tracked, however. So, our simulations are effectively ``2.5D'', i.e., 2D in physical space but 3D in momentum space.

We use a rectangular simulation box in the $xy$ plane, with periodic boundary conditions in the $y$ direction (\fig{simplane}). Each computational cell is initialized with two electrons and two ions, but we also performed limited experiments with a larger number of particles per cell (up to 8 per species), obtaining essentially the same results. The relativistic \tit{electron} skin depth for the incoming plasma ($c/\omega_{\rm pe}$)  is resolved with 10 computational cells and the simulation timestep is $\Delta t=0.045\,\omega_{\rm pe}^{-1}$. Here, $\omega_{\rm pe}\equiv(4\pi e^2 n_{e} /\gamma_0 m_e)^{1/2}$ is the relativistic electron plasma frequency for the upstream flow, with electron number density $n_{e}$ (measured in the wall frame) and bulk Lorentz factor $\gamma_0$. We employ a reduced mass ratio $m_i/m_e=16$, which allows to follow the shock evolution for longer times (in units of $\omega_{\rm pi}^{-1}=(m_i/m_e)^{1/2}\omega_{\rm pe}^{-1}$), while still clearly separating the ion and electron dynamical scales. As we show in Appendix \ref{sec:specmime}, we obtain essentially the same results when using higher mass ratios (we tried $m_i/m_e=100$, and up to $m_i/m_e=1000$ in some cases), which suggests that a mass ratio $m_i/m_e=16$ is already ``large'' enough to capture the correct acceleration physics in our shocks. In units of the relativistic \tit{ion} skin depth $\comp=(m_i/m_e)^{1/2}c/\omega_{\rm pe}$, the computational domain is $\sim25\comp$ wide (along $y$), which corresponds to 1024 cells for $m_i/m_e=16$ and 3072 cells for $m_i/m_e=100$. We also tried larger domains, up to 4096 cells wide (or $\sim100\comp$ for $m_i/m_e=16$), obtaining similar results. Our computational box expands along $x$ at the speed of light (see below), reaching the length of $\sim10,000\,\comp$ (or  $\sim400,000$ cells for $m_i/m_e=16$) at the final time of our longest simulation.

The incoming electron-ion stream is injected along $-\mathbf{\hat{x}}$ with bulk Lorentz factor $\gamma_0=15$ and a small thermal spread $\Delta\gamma=10^{-4}$. As we comment in \S\ref{sec:survey}, our results can be rescaled to account for a different $\gamma_0$ (we explore a wide range in $\gamma_0$, from 3 to 50). The upstream flow is seeded with a background magnetic field $B_0$ such that the ratio of magnetic to kinetic energy density is $\sigma\equiv B_{0}^2/4 \pi \gamma_0 m_i n_{i} c^2=0.1$, where $n_i$ (=$n_e$) is the number density of incoming ions.\footnote{The magnetization parameter as defined above pertains to ions. For electrons, $\sigma_{e0}=(m_i/m_e)\,\sigma$ at injection, where the electron kinetic energy is $\gamma_0m_ec^2$. So, it seems that the electron magnetization would depend on $m_i/m_e$, for fixed $\sigma$. However, on their way to the shock, electrons increase their average energy up to a fraction $\alpha$ of the initial ion energy $\gamma_0 m_i c^2$, where $\alpha$ is independent of $m_i/m_e$ (see Appendix \ref{sec:specmime}). So, the ``effective'' electron  magnetization $\sigma_{e,\rm{eff}}=(1/\alpha)\,\sigma$ does not depend on the mass ratio.}
In \S\ref{sec:survey} we show how our results change for lower magnetizations, down to $\sigma=10^{-5}$, at which point the shock properties approach the unmagnetized case discussed by \citet{spitkovsky_08}.  The magnetized regime we explore here may be relevant for internal shocks in GRBs and AGN jets, as we discuss in \S \ref{sec:disc}. 

For each value of the magnetization, we explore a range of magnetic inclinations by varying the angle $\theta$ between the shock direction of propagation $+\mathbf{\hat{x}}$ and the upstream magnetic field $\mathbf{B}_0$ (\fig{simplane}). The magnetic obliquity angle $\theta$ is measured in the wall frame. We vary $\theta$ from $\theta=0^\circ$, which corresponds to a ``parallel'' shock, with magnetic field aligned with the shock normal, up to $\theta=90^\circ$, i.e., a ``perpendicular'' shock, with magnetic field along the shock front. For $\theta\neq 0^\circ$, in the upstream medium we also initialize a motional electric field $\mathbf{E}_0=-\mbox{\boldmath{$\beta$}}_0\cross\mathbf{B}_0$, where  $\mbox{\boldmath{$\beta$}}_0=-\beta_0\;\bf{\hat{x}}$ is the three-velocity of the injected plasma. As we show in Appendix \ref{sec:specphi}, our results do not depend on the orientation of the magnetic field with respect to the simulation plane, as parameterized by the azimuthal angle $\varphi$ ($\varphi=0\deg$ for field in the $xy$ plane of the simulations, $\varphi=90\deg$ for field in the $xz$ plane).

As a byproduct of the shock evolution, particles and electromagnetic waves may propagate upstream from the shock at the speed of light. To ensure that the long-term evolution of the shock is captured correctly, they should not be removed from the computational domain, otherwise we would artificially suppress any feedback they may have on the shock. For this reason, we employ a ``moving injector'' (receding from the wall at the speed of light) and an expanding simulation box (see SS09 for details). This permits us to follow the shock evolution as far as the computational resources allow, preserving all the particles and waves generated by the shock. 

However, since the shock velocity is smaller than the speed of light, the distance between the injector and the shock increases linearly with time. In all PIC codes, a numerical heating instability arises when cold relativistic plasma propagates for large distances over the numerical grid \citep{dieckmann_06}. For \tit{electron-positron} shocks, in order to suppress this instability we let our moving injector periodically jump backward (i.e., towards the wall), so that the distance between the shock and the injector stayed roughly constant (see the ``jumping injector'' technique in SS09). However, for \tit{electron-ion} shocks this is not a viable solution, since the incoming flow is heavily influenced by particles and waves generated by the shock, as we show in \S\ref{sec:fluid}. By resetting the position of the injector, we would discard such particles and waves, thus dramatically interfering with the long-term evolution of the shock. Instead, we notice that the onset of the numerical instability is significantly delayed if we reduce the transverse size of our simulations. For long ($\gtrsim2000\omp$) simulations of strongly magnetized high-obliquity shocks, which are more affected by the numerical instability, we then employ a computational box with only 256 transverse cells (down to 64, in some cases). At $\ompt\lesssim2000$ (i.e., before the growth of the instability), the results from such small simulations are identical to our ``fiducial'' runs with 1024 transverse cells, showing that the relevant 2D properties of the shock are still preserved. For $m_i/m_e=16$, a transition to the 1D regime is observed only for computational boxes with less than 32 transverse cells.

%%%%%%%%%%%%%%%%%%%%%%%%%%%%%%%%%%%%%%%%%
\section{Shock Structure}\label{sec:fluid}
We now describe the structure and internal physics of relativistic magnetized electron-ion shocks. In this section, as well as in \S\ref{sec:accel}, we fix the upstream bulk Lorentz factor ($\gamma_0=15$) and magnetization ($\sigma=0.1$), and we investigate two representative magnetic inclination angles: $\theta=15\deg$ and $\theta=75\deg$. The  dependence of our findings on the bulk Lorentz factor, magnetization and magnetic obliquity of the upstream flow is discussed in \S\ref{sec:survey}. 

According to the criterion anticipated in \S\ref{sec:intro}, a shock is ``superluminal'' if particles cannot escape ahead of the shock by sliding along the magnetic field. In the upstream frame, this corresponds to magnetic inclinations $\theta'$ such that $\cos\theta'>\cos\theta'_{\rm crit}=\beta'_{\rm sh}$, where $\beta'_{\rm sh}$ is the  shock speed in the upstream frame. In the simulation frame, the critical obliquity angle that separates subluminal and superluminal configurations will be $\thetacrit={\rm arccot}[\gamma_{\rm sh}\,(\beta_0+\beta_{\rm sh})]$, where $\beta_{\rm sh}$ and $\gamma_{\rm sh}$ are the shock velocity and Lorentz factor in the simulation frame (see SS09 for a detailed discussion). For $\gamma_0=15$ and $\sigma=0.1$, the critical obliquity angle is $\theta_{\rm{crit}}\simeq 34^\circ$, and it stays confined within a relatively narrow range (between $\sim26\deg$ and $\sim42\deg$) for relativistic ($\gamma_0\gtrsim2$) flows with moderate magnetization ($\sigma\lesssim1.0$), as shown in Fig.~2 of SS09. It follows that $\theta=15\deg$ is a  subluminal shock, and $\theta=75\deg$ is superluminal.

\begin{figure}[tbp]
\begin{center}
\includegraphics[width=0.5\textwidth]{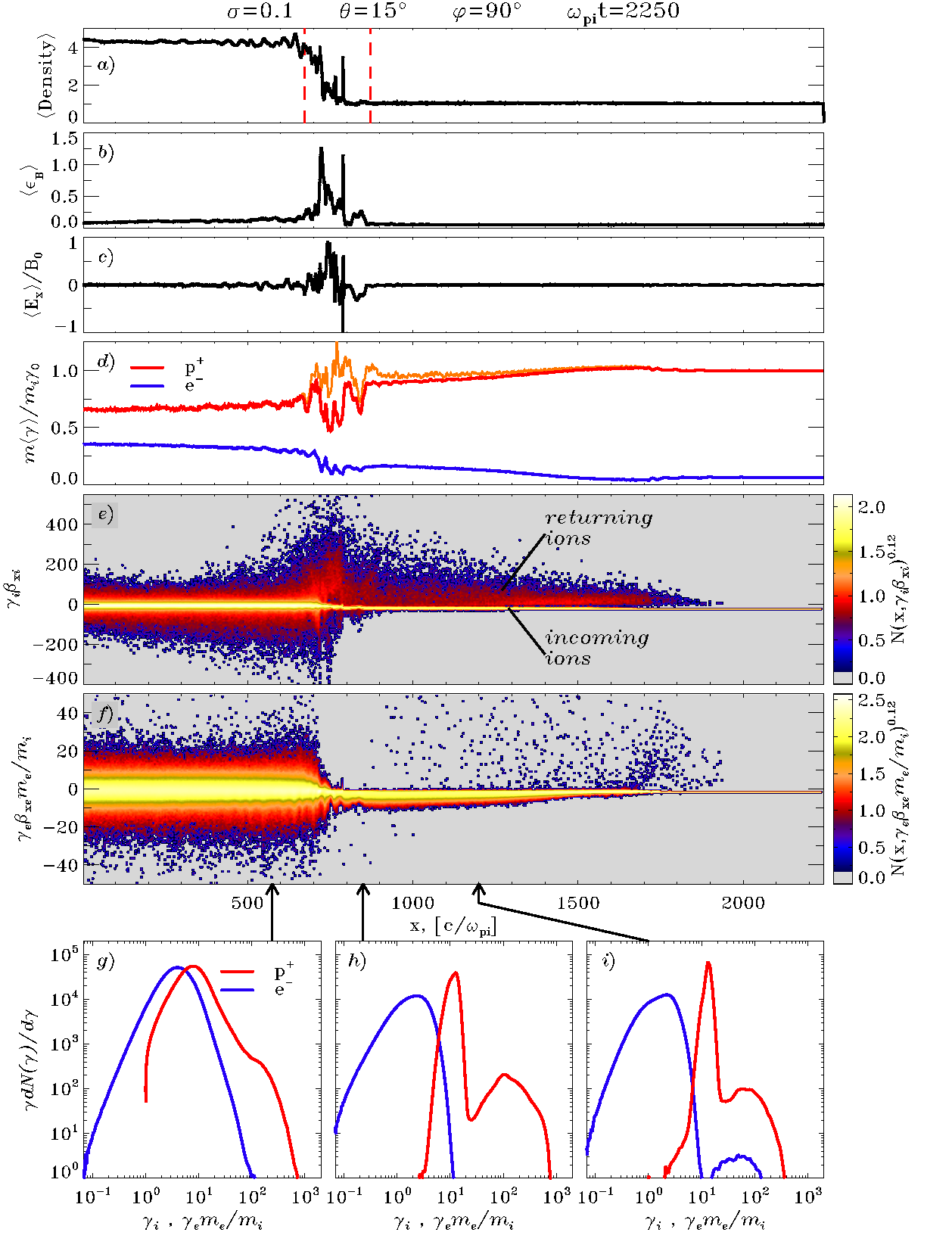}
\caption{Internal structure of a $\theta=15\deg$ subluminal shock at $\ompt=2250$. As a function of the longitudinal coordinate $x$, the following quantities are plotted: (a) $y$-averaged particle number density, in units of the upstream value; (b) $y$-averaged magnetic energy fraction $\epsilon_{\msc{b}}\equiv B^2/8 \pi \gamma_0 m_i n_{i} c^2$; (c) $y$-averaged longitudinal electric field $E_{\rm x}$, normalized to the upstream magnetic field $B_{0}$; (d) mean particle energy (red for ions, blue for electrons) for particles moving toward the shock, in units of the bulk energy of injected ions (orange line includes also the reflected ions; the equivalent for electrons overlaps with the blue line); longitudinal phase space of ions (e) and electrons (f); (g)-(i) particle energy spectra (red for ions, blue for electrons), at three locations across the flow, as marked by arrows at the bottom of panel (f).}
\label{fig:fluid15a}
\end{center}
\end{figure}

\begin{figure}[tbp]
\begin{center}
\includegraphics[width=0.5\textwidth]{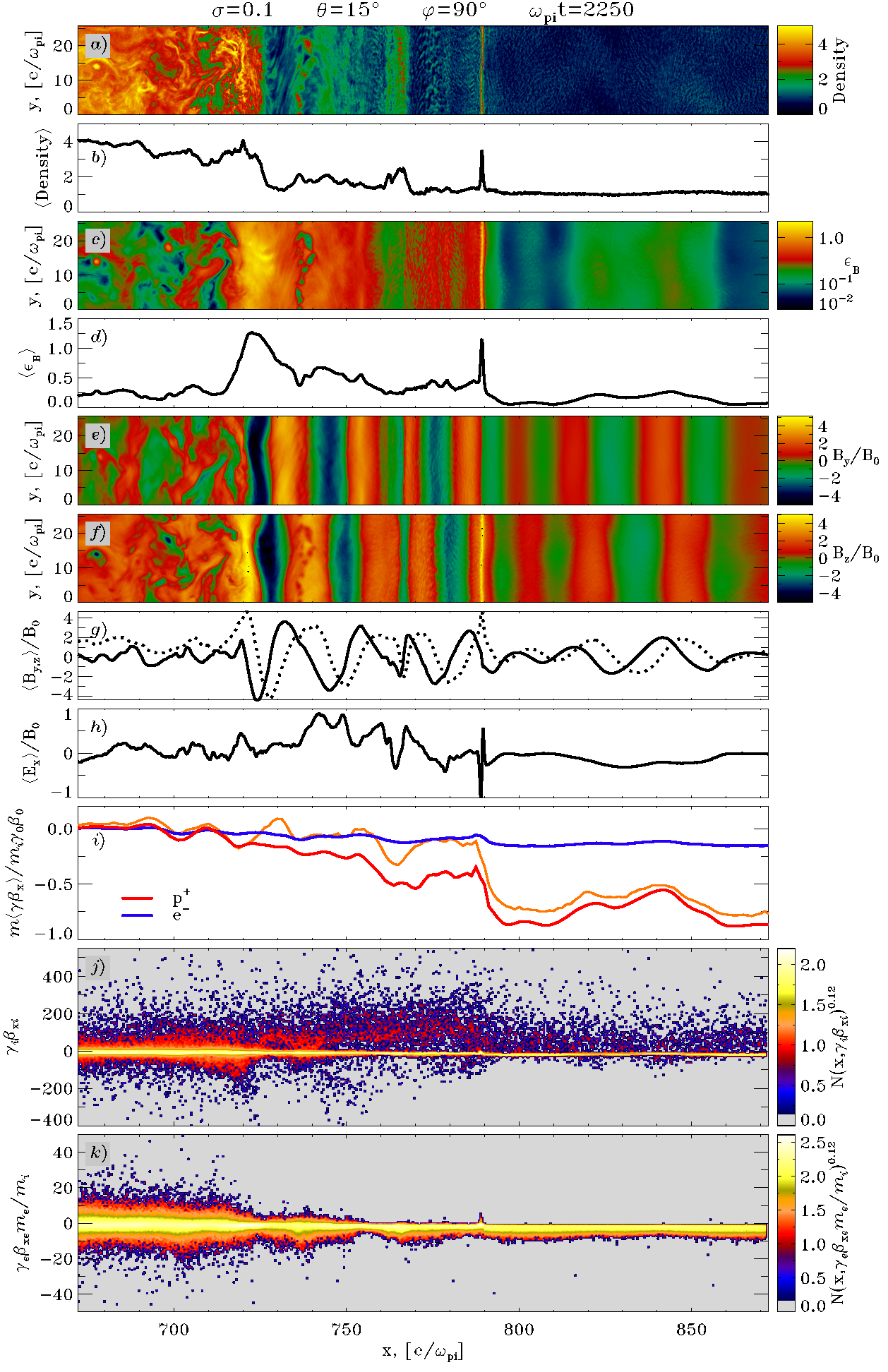}
\caption{Internal structure of a $\theta=15\deg$ subluminal shock at $\ompt=2250$, zooming in on a region around the shock, as delimited by the vertical dashed red lines in \fig{fluid15a}(a). The shock is located at $x\sim730\comp$. As a function of the longitudinal coordinate $x$, the following quantities are plotted: particle number density, in units of the upstream value (2D plot in the $xy$ simulation plane, panel (a); 1D $y$-averaged profile, panel (b)); magnetic energy fraction $\epsilon_{\msc{b}}\equiv B^2/8 \pi \gamma_0 m_i n_{i} c^2$ (2D in (c), 1D in (d)); 2D plots of $B_{\rm y}$ (e) and $B_{\rm z}$ (f), normalized to the upstream magnetic field $B_{0}$; (g) corresponding 1D profiles, for $B_{\rm y}$ (solid) and $B_{\rm z}$ (dashed); (h) 1D profile of the longitudinal electric field $E_{\rm x}$, in units of $B_{0}$; (i) mean particle $x$-momentum (red for ions, blue for electrons) for particles in the bulk (i.e., neglecting accelerated particles), in units of the momentum of injected ions (orange line includes also the accelerated ions; the equivalent for electrons overlaps with the blue line); longitudinal phase space of ions (j) and electrons (k).}
\label{fig:fluid15b}
\end{center}
\end{figure}

Figs.~\fign{fluid15a}-\fign{fluid75b} present the internal structure of the shock as a function of the longitudinal coordinate $x$. For the subluminal angle $\theta=15\deg$, \fig{fluid15a} covers the whole longitudinal extent of the simulation domain, whereas \fig{fluid15b} focuses on a smaller region around the shock, as delimited by the vertical dashed red lines in the first panel of \fig{fluid15a}. Both figures refer to $\ompt=2250$, when the shock is already fully developed. The temporal evolution of the shock, from the early stages up to the self-similar state approached at $\ompt\gtrsim2000$, is shown in \fig{denstime15}. Figs.~\fign{fluid75a} and \fign{fluid75b} correspond respectively to Figs.~\fign{fluid15a} and \fign{fluid15b} for the superluminal angle $\theta=75\deg$, at time $\ompt=1350$. Here, we choose an earlier time (with respect to $\ompt=2250$ for $\theta=15\deg$), to avoid the numerical instability discussed in \S\ref{sec:setup}, which grows at $\ompt\sim2000$ in high-obliquity shocks. Yet, time $\ompt=1350$ already captures the main properties of the long-term evolution of the shock.

This section is organized as follows. In \S\ref{sec:fover} we present a general description of the  structure of relativistic magnetized shocks in electron-ion plasmas. We focus on the properties they share with electron-positron shocks, and emphasize the differences. Then, we separately discuss the cases of subluminal (in \S\ref{sec:fsub}, for $\theta=15\deg$) and superluminal (in \S\ref{sec:fsuper}, for $\theta=75\deg$) shocks, describing how the obliquity of the field affects the shock structure.

%%%%%%%%%%%%%%%%%%%%%%%%%%%%%%%%%%%%%%%%%%%%%%
\subsection{Magnetized Relativistic Electron-Ion Shocks: General Overview}\label{sec:fover}
For all oblique magnetic configurations (i.e, with the exception of strictly parallel and perpendicular shocks), a fluid structure with two shocks (a strong ``fast'' shock and a weak ``slow'' shock) is seen in our PIC simulations of both electron-positron (SS09) and electron-ion flows, as expected from analytic theory \citep{majorana_anile_87} and MHD calculations \citep{komissarov_03b}. The incoming fluid does not stop completely at the fast shock. The residual bulk velocity is larger than the local slow magnetosonic speed, and a slow shock is formed behind the fast shock (i.e., closer to the wall, in our simulations). The slow shock transition in our PIC simulations is not as sharp as MHD would predict, due to insufficient dissipation along the magnetic field. The slow shock becomes slower and weaker as $\theta$ approaches either $0^\circ$ or $90^\circ$, and relatively stronger and faster for intermediate obliquities. For oblique configurations, we will neglect the slow shock and refer to the fast shock simply as ``the shock''.

The jump in density and electromagnetic fields at the fast shock, as well as the shock velocity, are in agreement with MHD calculations, and with PIC simulations of electron-positron shocks (see Table 1 in SS09). As observed in pair shocks, at low obliquities ($\theta\lesssim45\deg$) the effective adiabatic index of the downstream plasma is 4/3, as in a 3D relativistic gas, whereas for $\theta\gtrsim45\deg$ it tends to 3/2, as in a 2D fluid. In fact, right behind the shock the particle motion  is mostly confined to the plane orthogonal to the field, which for large inclination angles is nearly degenerate with the direction of propagation of the incoming flow. This prevents efficient isotropization along the field and results in a 2D adiabatic index.\footnote{We remark that this effect is not an artificial consequence of the reduced dimensionality of our computational domain, but it holds also for 3D simulations. Yet, it is more severe for 2D simulations with out-of-plane fields, since in this case the plane perpendicular to the field almost coincides with the simulation plane, and particle isotropization is even less efficient (see Appendix \ref{sec:specphi}).} Isotropization proceeds with distance behind the shock, and the particle distribution far downstream is roughly isotropic.

The mechanism that mediates randomization of particles at the shock varies depending on the field obliquity. In \tit{electron-positron} magnetized plasmas, SS09 found that low-obliquity shocks are mediated by Weibel-like filamentation instabilities \citep{weibel_59, medvedev_loeb_99, gruzinov_waxman_99}, as happens for unmagnetized shocks \citep[e.g.,][]{spitkovsky_05}.\footnote{More precisely, the instability results from the coupling of the electromagnetic filamentation (Weibel) instability with the electrostatic two-stream instability, as explained by \citet{bret_09}. The resulting modes are oblique with respect to the streaming direction.} 
The free energy for the instability comes from the counter-streaming between the incoming flow and the shock-accelerated particles propagating upstream. 

Low-obliquity shocks in \tit{electron-ion} magnetized plasmas are still mediated by counter-streaming instabilities, but the nature of the instability is different. For our fiducial values $\gamma_0=15$ and $\sigma=0.1$, the shock evolution at early times ($\ompt\lesssim300$) is governed by the \tit{electron} Weibel instability, which has the largest growth rate ($\sim\omega_{\rm pe}\gg\omega_{\rm pi}$). Later on, the so-called Bell's current-driven instability develops \citep{bell_lucek_01,bell_04,reville_06}, and the shock results from nonlinear steepening of the circularly-polarized Alfv\'enic-type waves associated with the instability (see \S\ref{sec:fsub} for details).  The source for the instability is the electric current of shock-accelerated ions that propagate ahead of the shock, and the polarization of the resulting modes is \tit{non-resonant} with respect to the ion gyro-motion. As we argue in Appendix \ref{sec:comparison},  Bell's instability governs the evolution of the shock only for a limited range of magnetizations ($10^{-2}\lesssim\sigma\lesssim0.3$, for fixed $\gamma_0=15$). If $\sigma\gtrsim0.3$, the polarization of the dominant mode changes, and the waves are now generated via gyro-frequency \tit{resonance} with the high-energy ions heading upstream \citep[][]{kulsrud_69}. For $\sigma\lesssim10^{-2}$, the magnetic field is dynamically unimportant, and the shock is mediated by the \tit{ion} Weibel instability \citep[as in unmagnetized shocks, see][]{spitkovsky_08}.

In contrast, high-obliquity shocks are mediated by magnetic reflection of the incoming flow off the shock-compressed magnetic field \citep[SS09]{alsop_arons_88}, for both electron-positron and electron-ion plasmas. The coherent gyration of the incoming particles at the shock triggers the so-called synchrotron maser instability \citep{hoshino_91}, where the ring of gyrating particles in the shock transition breaks up into bunches of charge. The bunches radiate a coherent train of transverse electromagnetic waves propagating into the upstream \citep{gallant_92, hoshino_92, spitkovsky_05}, with wavelength comparable to the Larmor radius of incoming \tit{electrons} in the shock-compressed fields. For \tit{electron-positron} shocks, this ``precursor wave'' does not have a significant impact on the upstream plasma (SS09). In \tit{electron-ion} flows, the guiding-center velocity of the incoming electrons decreases, since they experience relativistic transverse oscillations in the strong field of the  precursor wave. Due to their high mass, ions are less affected by the wave, and proceed at close to their initial velocity. The resulting difference in bulk velocity between the two species will generate a longitudinal electric field \citep{lyubarsky_06}, so that electrons are boosted toward the shock whereas ions are deboosted.\footnote{A clarification is required here regarding our terminology: by ``boost" or ``deboost'', we mean changes in the bulk $x$-momentum of the fluid; ``heating'' is used when the comoving particle distribution broadens; finally, ``acceleration'' is something pertaining only to a subsample of the particles, extracted from the bulk of the fluid and energized up to suprathermal velocities.}
As a result, energy equipartition between electrons and ions may be achieved even before the flow arrives at the shock front, as we discuss in \S\ref{sec:fsuper}.

On the other hand, if the efficiency of ion-to-electron energy transfer \tit{in the upstream} is moderate or negligible, the incoming ions will enter the shock with bulk energy much larger than electrons, so that they will penetrate deeper into the shock-compressed field (roughly, in proportion to the ratio of ion to electron Larmor radii). The resulting charge separation establishes a net electrostatic field  \tit{at the shock} pointing toward the upstream. The associated cross-shock electric potential attracts the incoming electrons into the downstream and decelerates the incoming ions, so that electrons may still end up with a significant fraction of the ion energy.

%%%%%%%%%%%%%%%%%%%%%%%%%%%%%%%%%%%%%%%%%%%%%%%%%%%%

\subsection{Subluminal Shocks: $0\deg\leq\theta<\thetacrit$}\label{sec:fsub}
In this section, we describe the internal structure of subluminal magnetized electron-ion shocks. We take $\theta=15\deg$ as a representative case, but our results will apply to the whole range of subluminal angles, unless otherwise noted. In both electron-positron and electron-ion plasmas, subluminal shocks are characterized by the presence of a diffuse stream of shock-reflected particles that propagate ahead of the shock. The counter-streaming between these ``returning'' particles and the incoming flow triggers the generation of waves in the upstream medium, which in turn affect the structure of the shock and the process of particle acceleration.

The longitudinal ion phase space in \fig{fluid15a}(e) shows the injected ions as a cold dense beam with $\gbi{x}\simeq-15$. In the shock transition layer (located at $\sim730\comp$ for $\ompt=2250$), the incoming ions are isotropized and thermalized. A diffuse population of high-energy ions moves ahead of the shock with $\gbi{x}>0$, following the upstream magnetic field; as we show in \S\ref{sec:asub_mech}, these ``returning'' ions have been accelerated at the shock. In the upstream ion spectrum (red lines in \fig{fluid15a}(h) and (i)), they appear as a high-energy broad bump, whereas the low-energy narrow peak is populated by the incoming particles. If transmitted downstream, the shock-accelerated ions will populate a power-law nonthermal tail, which is seen in the energy spectrum of \fig{fluid15a}(g) (red line) beyond the thermal distribution. 

In contrast, very few electrons propagate back upstream from the shock (see the longitudinal phase space of electrons in \fig{fluid15a}(f)). This anticipates that in subluminal magnetized shocks electrons are accelerated with lower efficiency than ions.\footnote{The very few high-energy electrons seen with $\gamma_e\beta_{{\rm x}e}>0$ ahead of the shock (see \fig{fluid15a}(f), and the high-energy bump in the blue line of \fig{fluid15a}(i)) have been accelerated in the first stages of evolution, when the shock was mediated by the electron Weibel instability. When ion-driven instabilities take over, electron acceleration shuts off.}
Despite the lack of returning electrons, the electron spectrum downstream from the shock (blue line in panel (g)) deviates from a purely thermal distribution, which suggests that some electrons may be accelerated to suprathermal energies as they cross the shock from upstream to downstream. We refer to \S\ref{sec:asub} for a detailed discussion of electron acceleration.
 
The fact that returning ions greatly outnumber electrons has multiple effects on the structure of the shock, which were absent in electron-positron flows. First, the positively-charged cloud of returning ions perturbs the incoming flow, causing significant transfer of energy from ions to electrons. At the leading edge of the population of returning ions ($x\sim1800\comp$), the incoming particles cross a kind of double layer, where ions are boosted toward the shock and electrons are deboosted (see the average particle energy in \fig{fluid15a}(d), at $x\sim1700\comp$; red for ions, blue for electrons). Beyond that point, electrons are significantly heated (see \fig{fluid15a}(f) at $800\comp\lesssim x\lesssim1500\comp$), possibly via small-scale plasma oscillations (Spitkovsky et al., in prep). The average electron energy increases toward the shock at the expense of the ion bulk energy, reaching up to $\sim15\%$ of the kinetic energy of injected ions (blue line in \fig{fluid15a}(d)). However, electrons still enter the shock with lower average energy than ions. The resulting charge separation in the shock-compressed magnetic field generates a net cross-shock electrostatic field ($E_{\rm x}>0$, seen at $x\sim730\comp$ in \fig{fluid15a}(c)). The associated cross-shock potential further increases the average electron energy; in the downstream, electrons have $\sim30\%$ of the initial ion energy.
 
Furthermore, the imbalance between returning ions and electrons establishes a net (ion-driven) current in the upstream medium. The compensating current by the incoming electrons triggers the growth of the so-called Bell's instability, which is expected in this regime of shock parameters \citep{bret_09, pelletier_10}. As shown in panel (e) (for $B_{\rm y}$) and (f) (for $B_{\rm z}$) of \fig{fluid15b}, the instability generates circularly-polarized Alfv\'enic-type waves in the upstream medium, with wavevector along the background field.\footnote{This is most clear in simulations with in-plane $\mathbf{B}_0$ (i.e., $\varphi=0\deg$). Here, $\varphi=90\deg$, and one can only appreciate that the projection of the wavevector onto the simulation plane lies along $\mathbf{\hat{x}}$ (see Figs.~\ref{fig:fluid15b}(e) and (f)).} The polarization of the waves is such that they are non-resonant with respect to the returning ions, and their wavelength ($\sim25\comp$) is much smaller than the Larmor radius of the highest-energy particles. As shown in \fig{fluid15b}(g) (solid for $B_{\rm y}$, dotted for $B_{\rm z}$), the amplitude of the waves increases as they approach the shock, reaching up to $\sim4$ times the strength of the initial field (measured in the simulation frame). Right before the shock ($x\sim730\comp$), more than $50\%$ of the bulk kinetic energy of the upstream flow has been converted into wave magnetic energy (see the magnetic energy fraction $\epsilon_{\ditto B}\equiv B^2/8 \pi \gamma_0 m_i n_{i} c^2$ in \fig{fluid15b}(d)). After the shock, the flow isotropizes, the electric current required to drive the instability vanishes, and the waves dissipate their energy into particle energy. The importance of Bell's waves for ion acceleration will be discussed in \S\ref{sec:asub_mech}.

In the vicinity of the shock ($730\comp\lesssim x\lesssim790\comp$), the pressure of returning ions is so large that they heavily impact the structure of the transition region and the dynamics of the incoming flow. Due to the push of the returning ions, the incoming plasma slows down significantly, by more than $50\%$, well before the shock front (see the jump in the average $x$-momentum of incoming ions at $x\sim790\comp$, solid red line in \fig{fluid15b}(i)). The deceleration and consequent compression of the incoming fluid produces spikes in the number density and the magnetic energy, as observed at $x\sim790\comp$ in \fig{fluid15b}(a)-(d). Such spikes are transient quasi-periodic structures, which get  advected downstream after $\sim100\omp$ (or $\sim30\,\omega_{\rm ci}^{-1}$, with the relativistic ion Larmor frequency $\omega_{\rm ci}\equiv\sqrt{\sigma}\,\omega_{\rm pi}$), and then reform at a slightly larger distance from the shock.\footnote{They resemble the so-called Short Large-Amplitude Magnetic Structures, or SLAMS, observed at the Earth's bow shock \citep[e.g.,][]{schwartz_92}.} This is the characteristic signature of the so-called ``shock reformation'' process. Evidence of shock modification by the accelerated particles may also be seen in the ion energy spectrum, as we discuss in \S\ref{sec:asub_time}.

\subsubsection{Time Evolution}

\begin{figure}[tbp]
\begin{center}
\includegraphics[width=0.5\textwidth]{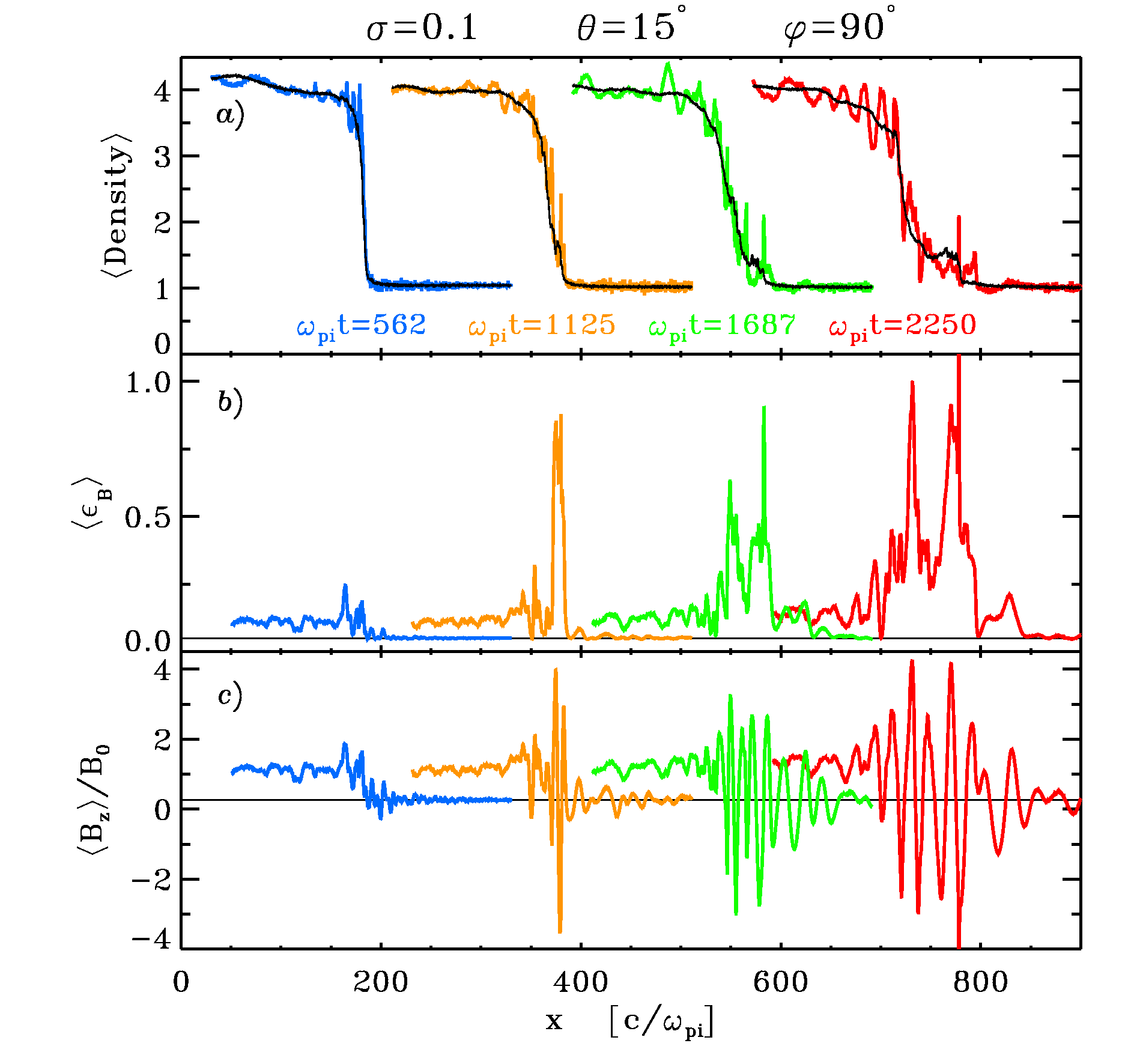}
\caption{Time evolution of the internal structure of a $\theta=15\deg$ subluminal shock: $\ompt=562$ (blue), $\ompt=1125$ (yellow), $\ompt=1687$ (green), and $\ompt=2250$ (red). (a) $y$-averaged profiles of the particle number density;  for each time, the superimposed thin black line is obtained by averaging over a narrow  temporal window ($100\,\omega_{\rm pi}^{{-1}}$ wide), centered on that time. (b) $y$-averaged profiles of the magnetic energy fraction $\epsilon_{\msc{b}}$, with the horizontal black line showing the value  at injection ($=0.5\,\sigma=0.05$). (c) $y$-averaged profiles of the transverse magnetic field $B_{\rm z}$ (in units of $B_0$), with the horizontal black line showing the value  at injection ($=\sin15\deg\simeq0.26$). For each time, we only plot a limited region ($\sim300\comp$ wide) around the instantaneous location of the shock.}
\label{fig:denstime15}
\end{center}
\end{figure}

The picture outlined above refers to $\ompt=2250$, when the number of shock-accelerated ions ahead of the shock is already large enough to affect the structure of the transition region. \fig{denstime15} follows the shock evolution from $\ompt=562$ to $\ompt=2250$, showing how the $y$-averaged profiles of number density (panel (a)), magnetic energy fraction $\epsilon_{\msc{b}}\equiv B^2/8 \pi \gamma_0 m_i n_{i} c^2$ (panel (b)), and transverse magnetic field $B_{\rm z}$ (panel (c)) change with time.

At early times ($\ompt=562$, blue line), when the pressure of returning ions is negligible with respect to the ram pressure of the incoming flow, the density jump occurs on the scale of a few ion Larmor radii, like an ideal MHD shock. The magnetic field at the shock ($\epsilon_{\msc{b}}\sim0.15$, blue line in panel (b)) mostly results from the compression of the upstream field, with a minor contribution from counter-streaming instabilities.

Starting from $\ompt=1125$ (yellow line), the shock structure changes. For $\theta=15\deg$, the leading edge of the population of shock-accelerated ions recedes from the shock with $x$-velocity $\simeq0.86\,c$ (corresponding to particles moving along the background field at the speed of light). Since the shock speed is $\simeq0.32\,c$, the distance between the shock and the head of returning ions increases with time. Bell's instability, which at earlier times was suppressed by advection into the shock, can now amplify the field to highly nonlinear values ($B_{\rm z}$ up to a factor of $\sim4$ larger than $B_0$, see green and red lines in panel (c); similar values for $B_{\rm y}$). The magnetic energy at the shock saturates at equipartition with the upstream kinetic energy ($\epsilon_{\msc{b}}\sim0.5$, green and red lines in panel (b)).

The increase in the strength of Bell's waves is driven by the growth of the population of returning ions. At late times (green for $\ompt=1687$, red for $\ompt=2250$), the shock transition region becomes much wider (panel (a)), due to the increasing push of the returning ions on the incoming flow. The time-averaged density profile (see the thin black line in panel (a) for $\ompt=2250$) displays the typical structure of ``cosmic-ray modified'' shocks \citep[e.g.,][]{berezhko_99}, with a smooth density increase (the ``cosmic-ray precursor'') ahead of the main shock (here properly called ``subshock''), now located at $x\sim730\comp$. The magnetic energy stays large ($\epsilon_{\msc{b}}\sim0.5$) across the whole transition region, with two characteristic peaks, one at the head of the cosmic-ray precursor, where the incoming plasma is first decelerated and compressed, and the other at the subshock, where the flow comes to rest (see green and red lines in panel (b)). As clarified by panel (c), most of the magnetic energy in the region between the peaks results from the compression of Bell's waves.

At later times (we follow the shock evolution up to $\ompt=5062$), the shock structure does not qualitatively change, but the thickness of the transition region (where $\epsilon_{\msc{b}}\sim0.5$) still expands at $\simeq0.1\,c$. A detailed study of the relation between shock width and mean free path of diffusively-accelerated ions will be presented elsewhere.

%%%%%%%%%%%%%%%%%%%%%%%%%%%%%%%%%%%%%%%%%%%%%%%%%%%%%%
\begin{figure}[tbp]
\begin{center}
\includegraphics[width=0.5\textwidth]{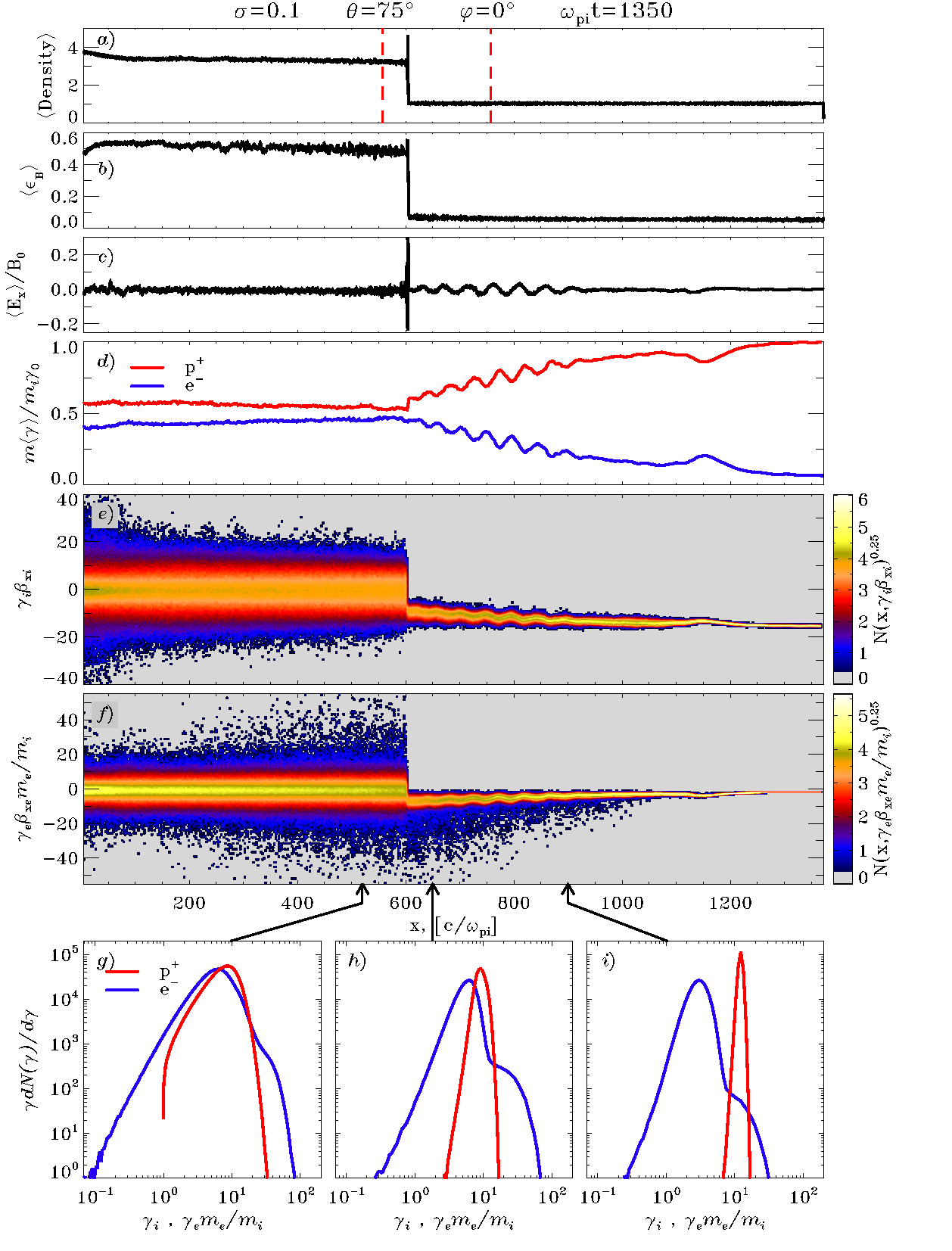}
\caption{Internal structure of a $\theta=75\deg$ superluminal shock at $\ompt=1350$. See the caption of \fig{fluid15a} for details. Here, the slow shock can be seen at $x\lesssim50\comp$ as an increase in density (panel (a)) and decrease in magnetic energy (panel (b)).}
\label{fig:fluid75a}
\end{center}
\end{figure}
\begin{figure}[tbp]
\begin{center}
\includegraphics[width=0.5\textwidth]{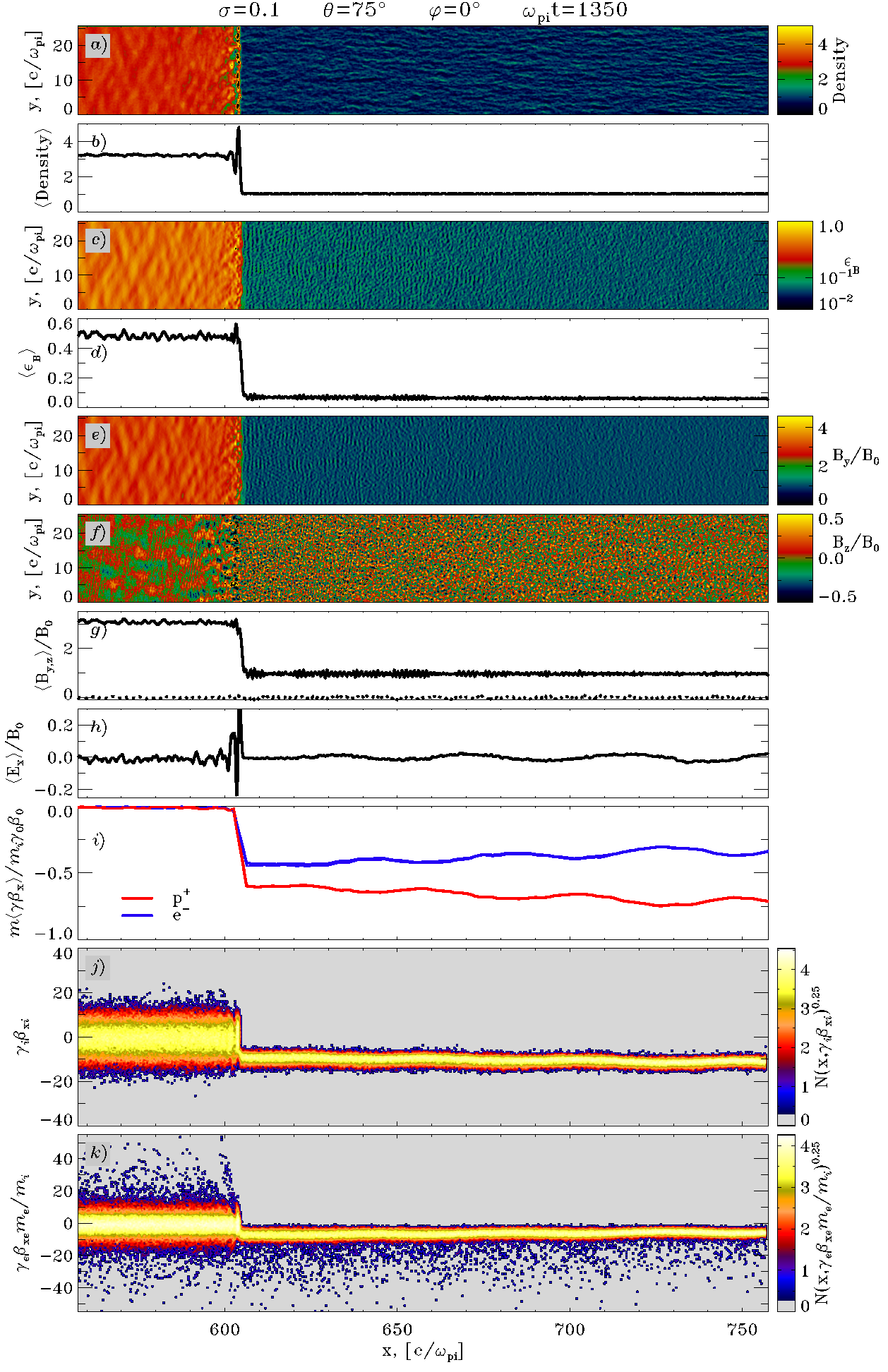}
\caption{Internal structure of a $\theta=75\deg$ superluminal shock at $\ompt=1350$, zooming in on a region around the shock, as delimited by the vertical dashed red lines in \fig{fluid75a}(a). See the caption of \fig{fluid15b} for details.}
\label{fig:fluid75b}
\end{center}
\end{figure}

\subsection{Superluminal Shocks: $\thetacrit<\theta\leq90\deg$}\label{sec:fsuper}
We now turn to the structure of a shock with obliquity $\theta=75\deg$, as a representative example of superluminal shocks. In such shocks, particles following the magnetic field cannot propagate back from the shock into the upstream.  In fact, in \fig{fluid75a} no particles are seen ahead of the shock with positive $x$-momentum (neither ions nor electrons, phase spaces in \fig{fluid75a}(e) and (f) respectively). The shock transition region (located at $x\sim600\comp$ for $\ompt=1350$), which is not perturbed by the pressure of energetic particles, is much thinner than for $\theta=15\deg$, on the scale of a few Larmor radii of the incoming ions (compare the density profile in panel (b) between \fig{fluid15b} and \fig{fluid75b}). Despite the absence of returning particles, the incoming flow can still be perturbed by the presence of the shock, e.g., via electromagnetic waves emitted by the shock into the upstream.

As discussed in \S\ref{sec:fover}, coherent Larmor gyration of the incoming electrons in the shock-compressed field triggers the synchrotron maser instability \citep{hoshino_91}, which creates a train of transverse electromagnetic waves propagating into the upstream. These ``precursor'' waves can be barely seen as short-scale ripples stretched along $y$ in the upstream region of \fig{fluid75b}(c) (for $\epsilon_{\ditto B}$) and  \fig{fluid75b}(e) (for $B_{\rm y}$),  especially in the vicinity of the shock.

The incoming electrons oscillate in the transverse field of the precursor waves, and their guiding-center velocity decreases. Since ions have larger inertia, their speed is not appreciably altered by the waves. Behind the head of the precursor ($x\sim1200\comp$ in \fig{fluid75a}), the difference between the bulk velocities of ions and electrons generates a longitudinal ``wakefield'' $E_{\rm x}>0$ \citep{lyubarsky_06}, so that electrons are initially boosted toward the shock, whereas ions are deboosted (see the average particle energy in \fig{fluid75a}(d) at $x\gtrsim1150\comp$; red for ions, blue for electrons). This initiates electrostatic wakefield oscillations in the incoming plasma (see $E_{\rm x}$  wiggles in \fig{fluid75a}(c) at $600\comp\lesssim x\lesssim900\comp$), that mediate a \tit{quasi-periodic} exchange of energy between ions and electrons in the upstream (see analogous wiggles in  \fig{fluid75a}(d)).

If the strength of the precursor were uniform throughout the upstream region, after each wakefield oscillation the electron Lorentz factor would come back to the value $\gamma_0$ at injection \citep{lyubarsky_06}. Instead, we observe a \tit{secular} increase in the average energy of electrons (and decrease for ions) as the incoming flow approaches the shock (\fig{fluid75a}(d) at $600\comp\lesssim x\lesssim1100\comp$). This corresponds to a gradient in the amplitude of the precursor waves, which are stronger in the vicinity of the shock. Their radiative push on the incoming electrons will then be larger right in front of the shock, and weaker farther upstream. This imbalance establishes a net wakefield $E_{\rm x}>0$ throughout the upstream region (in addition to the oscillations discussed above), which results in a systematic increase of the average electron energy toward the shock, as seen in  \fig{fluid75a}(d). Alternatively, one could define a ``ponderomotive'' electric potential (proportional to the radiative push of the precursor), whose increase toward the shock would explain the profile of average electron energy in \fig{fluid75a}(d)  \citep[][]{hoshino_08}.

As a result, electrons and ions enter the shock with roughly the same energy. Given the resulting absence of electron-ion charge separation at the shock, no significant cross-shock electric field is generated, so that the electron and ion average energies do not appreciably change across the shock (\fig{fluid75a}(d) at $x\sim600\comp$). So, in high-obliquity shocks the upstream ponderomotive potential  turns out to be more important than the cross-shock potential, yet it is not included in most models of relativistic magnetized shocks \citep[e.g.,][]{gedalin_08}. 

The increase in electron bulk energy toward the shock is accompanied by substantial electron heating, i.e., broadening of the electron distribution. In the upstream, the electron spectrum shows two components (see blue line in \fig{fluid75a}(h) and (i)). The low-energy ``thermal'' part, that contains most of the particles, broadens and shifts to higher energies as the flow propagates toward the shock (the shift corresponds to the secular increase in electron energy of \fig{fluid75a}(d)). The high-energy component, which gets more populated in the vicinity of the shock, extends a factor of $\sim10$ higher in energy than the thermal peak. It is seen in the electron phase space of \fig{fluid75a}(f) as a diffuse beam moving toward the shock, with much higher $x$-momentum than the bulk flow. In the downstream spectrum, this extra component survives as a tail at high energies (blue line in \fig{fluid75a}(g)). 

As we discuss in \S\ref{sec:asuper_time}, this high-energy component should be interpreted as a separate hotter (but still thermal) population, rather than as a nonthermal tail of accelerated particles. It is populated by electrons that, on their way to the shock, happen to receive a substantial kick toward the upstream by the wakefield oscillations mentioned above. At first, the energy of these electrons will decrease, because the kick is opposite to their $x$-momentum. As explained in  \ref{sec:asuper_mech}, the electron energy will then oscillate with the gyro-period, periodically reaching a maximum that, for strong kicks, may be much larger than the characteristic energy of electrons in the bulk (see \fig{fluid75a}(f)). In support of this interpretation, we see in the phase space of \fig{fluid75a}(f) that injection of electrons into the high-energy component starts at $x\sim1050\comp$, right behind the location where the average electron energy is driven to a minimum  (\fig{fluid75a}(d) at $x\sim1100\comp$)  by the wakefield oscillations in \fig{fluid75a}(c). Closer to the shock, where the wakefield oscillations are stronger (\fig{fluid75a}(c) at $600\comp\lesssim x\sim900\comp$), more electrons are kicked back upstream and injected into the high-energy spectral component. This explains why the electron high-energy tail gets more populated toward the shock (compare \fig{fluid75a}(h) and (i)). In the following, since the high-energy component in the electron spectrum is ultimately powered by the dissipation of wakefield oscillations, we shall refer to this process as ``wakefield heating''.

In contrast, ions are not significantly affected by the wakefield oscillations, due to their larger mass. The upstream ion spectrum is then consistent with a single component (red line in \fig{fluid75a}(h) and (i)), which results in a purely Maxwellian distribution in the downstream region (red line in \fig{fluid75a}(g)).

We remark that the main features discussed here for $\theta=75^\circ$, and in particular the emergence of a separate component of hot electrons in the upstream, are common to all superluminal configurations, up to $\theta=90^\circ$.

%%%%%%%%%%%%%%%%%%%%%%%%%%%%%%%%%%%%%%%%%%%%%%%%%
\section{Particle Acceleration}\label{sec:accel}
In this section, we discuss particle acceleration in subluminal and superluminal shocks. We fix the magnetization ($\sigma=0.1$) and bulk Lorentz factor ($\gamma_0=15$) of the upstream flow, and we investigate one representative subluminal ($\theta=15\deg$, in \S\ref{sec:asub}) and one superluminal ($\theta=75\deg$, in \S\ref{sec:asuper}) magnetic obliquity. For each case, we describe the time evolution of the downstream energy spectrum and the mechanism responsible for particle energization.

%%%%%%%%%%%%%%%%%%%%%%%%%%%%%%%%%%%%%%

%%%%%%%%%%%%%%%%%%%%%%%%%%%%%%%%%%%%%%
\subsection{Subluminal Shocks: $0\deg\leq\theta<\thetacrit$}\label{sec:asub}
\begin{figure}[tbp]
\begin{center}
\includegraphics[width=0.5\textwidth]{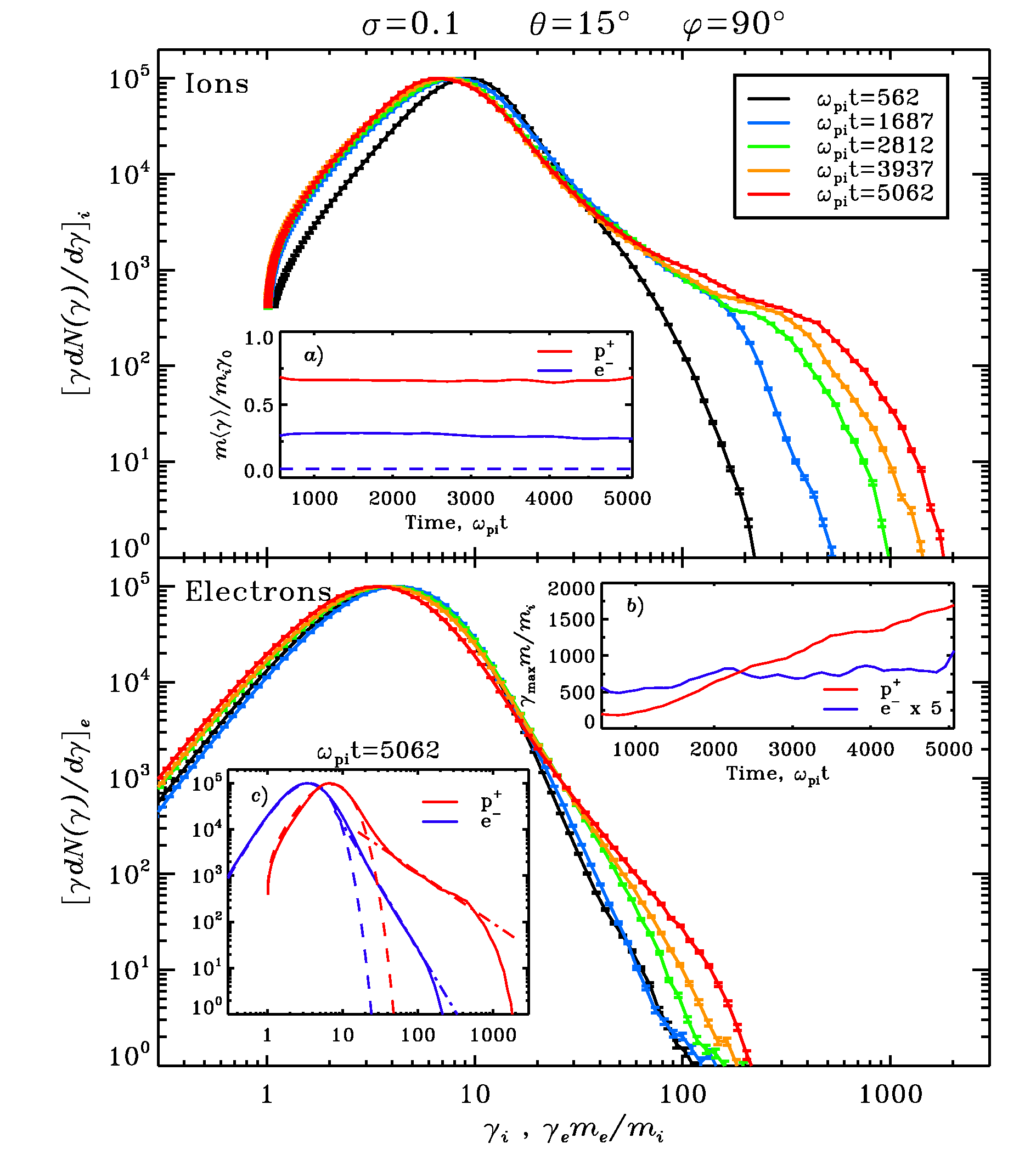}
\caption{Time evolution of the downstream energy spectrum in a $\theta=15\deg$ subluminal shock, for ions (upper panel) and electrons (lower panel): $\ompt=562$ (black), $\ompt=1687$ (blue), $\ompt=2812$ (green), $\ompt=3937$ (yellow), and $\ompt=5062$ (red). Subpanels: (a) time evolution of the downstream \tit{mean} particle energy (red for ions, blue for electrons), in units of the bulk energy of injected ions, with the horizontal dashed blue line showing the value expected for electrons in the absence of any ion-to-electron energy transfer ($=m_e/m_i\simeq0.06$); (b) time evolution of the downstream \tit{maximum} particle energy (red for ions; blue for electrons, multiplied by 5 for clarity); (c) fit to the electron (blue) and ion (red) spectrum at $\ompt=5062$, with a low-energy Maxwellian (dashed) plus a high-energy power law (dot-dashed). In panel (c), axes are the same as in the main plot.}
\label{fig:spectime15}
\end{center}
\end{figure}

\subsubsection{Time Evolution}\label{sec:asub_time}
\fig{spectime15} shows the time evolution of the ion (upper panel) and electron (lower panel) energy spectrum for $\theta=15\deg$, in a downstream slab at fixed distance from the shock. As shown in panel (c), the spectrum of both ions (red) and electrons (blue) clearly deviates from a Maxwellian (dashed), with a substantial population of nonthermal particles arranged in a power-law tail (dot-dashed).

The high-energy tail in the ion spectrum (upper panel)  grows with time as more and more particles are shock-accelerated. At late times (red curve for $\ompt=5062$), the slope approaches $-2$, which corresponds to equal energy contributions by each decade in Lorentz factor. It follows that, as the upper cutoff of the spectrum increases linearly in time (red line in panel (b)), the energy content of the ion nonthermal tail steadily grows, reaching $\sim30\%$ of the total ion energy at $\ompt=5062$.  By number, the tail is still dominated by the lower energies, and its fractional contribution to the ion census saturates at $\sim5\%$. At late times (e.g., $\ompt=5062$), the tail does not resemble a simple power-law with constant slope, but it seems to be somewhat concave, which may be due to the modification of the shock by accelerated ions (see \S\ref{sec:fsub}), as discussed for non-relativistic shocks in SNRs by \citet{amato_blasi_05, amato_blasi_06}. 

As more ion energy is stored in the nonthermal tail, the peak of the ion thermal distribution shifts to lower energies. A similar trend is observed for electrons (lower panel in \fig{spectime15}), with the thermal bump moving to lower Lorentz factors as the acceleration efficiency increases.\footnote{We note that, whereas the partition of energy between the thermal peak and the nonthermal tail changes with time, the average electron energy (that includes both components) stays relatively constant, at $\sim30\%$ of the initial ion energy (blue line in \fig{spectime15}(a)).}
However, the tail in the electron spectrum contains less particles ($\sim2\%$) and energy ($\sim10\%$) than the ion tail, it is much steeper (with slope $-3.5\pm0.1$ at $\ompt=5062$), and its upper energy cutoff  does not grow very fast in time (blue line in panel (b), multiplied by 5 for clarity). Most importantly, as discussed in \S\ref{sec:fsub}, no electrons are seen to propagate back upstream from the shock, as opposed to the large population of returning ions (compare the phase spaces in panels (e) and (f) of \fig{fluid15a}). This suggests that there should be a substantial difference between the acceleration process of ions and electrons, as we now describe.

\subsubsection{Energization Mechanism}\label{sec:asub_mech}

\begin{figure}[tbp]
\begin{center}
\includegraphics[width=0.5\textwidth]{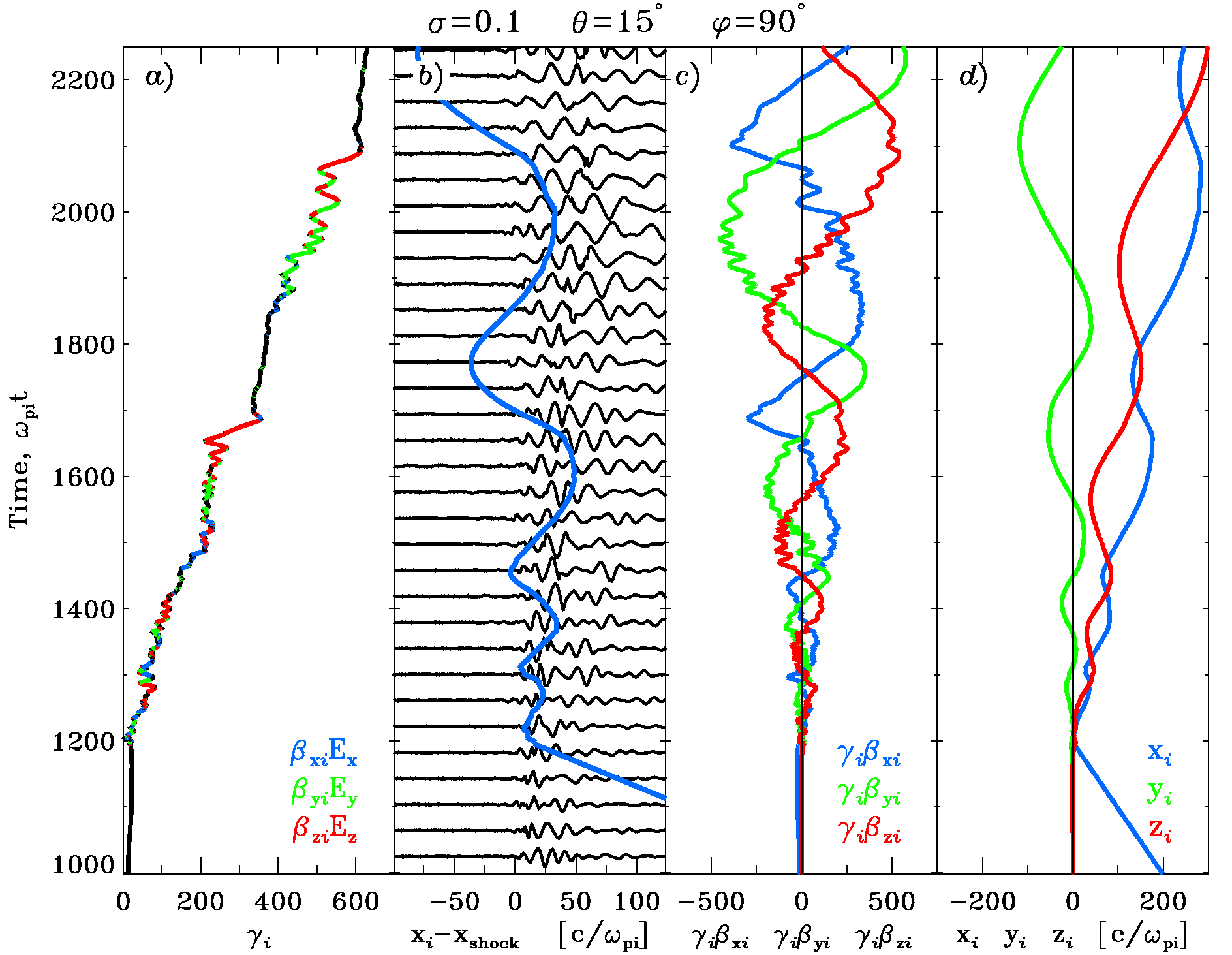}
\caption{Trajectory of a representative high-energy ion extracted from the simulation of a $\theta=15\deg$ subluminal shock. Panel (a): time evolution of the particle energy; different colors show which component of the electric field governs the energy change (blue for $E_{\rm x}$, green for $E_{\rm y}$, red for $E_{\rm z}$), for the portions of the trajectory with the largest rate of acceleration or deceleration. Panel (b): $x$-location of the particle relative to the shock, superimposed over $y$-averaged profiles of $E_{\rm z}$ stacked in time. The electric field is still measured in the wall frame, but shifted along $x$ so that the shock appears stationary. Panel (c): time evolution of the particle 4-velocity, with the same color coding as in panel (a). Panel (d): time evolution of the particle position ($x$-location is relative to the first encounter with the shock), with the same color coding as in panel (a). We remind that in this case ($\varphi=90\deg$) the upstream background electric field $\mathbf{E}_0$ is along $-\mathbf{\hat{y}}$.}
\label{fig:ions15}
\end{center}
\end{figure}

The orbit of a representative high-energy ion from the simulation of a subluminal shock ($\theta=15\deg$) is plotted in \fig{ions15}. For $\ompt\lesssim1200$, the particle is approaching the shock from upstream with bulk $x$-momentum $\sim-\gamma_0\beta_0\simeq-15$. From $\ompt\sim1200$ to $\ompt\sim2100$, it stays at the shock gyrating around the mean magnetic field  (see the ion $x$-location relative to the shock in panel (b)), and its Lorentz factor increases up to $\gamma_i\sim600$ (panel (a)). Energy gain occurs primarily when the ion is in the upstream region, whereas its Lorentz factor does not significantly change while downstream (e.g., from $\ompt\sim1700$ to $\ompt\sim1850$).\footnote{This is a consequence of our choice for the simulation frame, which coincides with the downstream plasma frame. It follows that, in the downstream region, no motional electric field is present, and the particle energy does not appreciably change. In the shock frame, energy change would be seen on both sides of the shock.}
 As the particle energy grows, its Larmor radius also proportionally increases (see the particle trajectory in panel (d)). The selected ion is finally transmitted downstream at $\ompt\sim2100$, where it will populate the nonthermal tail seen in the ion spectrum of \fig{spectime15}. In contrast, the shock-accelerated ions that are eventually reflected back upstream will contribute to the beam of returning ions seen in \fig{fluid15a}(e), which are responsible for triggering the growth of Bell's instability (see \S\ref{sec:fsub}). In turn, the waves generated by the instability mediate the process of ion acceleration, as we now describe.
 
In panel (a), different colors show which component of the electric field governs the energy change. We see that energy gain is primarily associated with $E_{\rm y}$ (green) and $E_{\rm z}$ (red), which are indeed the two components associated with Bell's circularly-polarized waves (see \fig{fluid15b}(g)). From the 1D profiles of $E_{\rm z}$ time-stacked in panel (b), we see that the wavelength of Bell's modes is typically smaller than the ion Larmor radius (compare with blue line in panel (b)). It follows that the interaction between the selected ion and the waves will be \tit{non-resonant}, i.e., during the half Larmor period when the ion is in the upstream, it will encounter multiple wave fronts, resulting in repeated accelerations and decelerations on a sub-Larmor scale. This is clearly shown in the energy evolution of panel (a) and in the 4-velocities of panel (c), where Bell's waves cause short-scale wiggles superimposed over the (otherwise smooth) Larmor gyration.  

The overall energy gain results from favorable encounters (i.e., with the ion velocity locally aligned with the wave electric field) being more frequent than unfavorable ones. In particular, significant energization may occur when the ion is crossing the shock from upstream to downstream, as seen at $\ompt\sim1675$ and $\ompt\sim2075$ (compare panels (a) and (b)).  In this case, the ion can remain in phase with the  waves for a larger fraction of its orbit, since both the ion and the waves are moving toward the shock. Efficient acceleration results if the wave electric field is aligned with the ion velocity (mostly oriented along $+\mathbf{\hat{z}}$ in this phase, for a positively-charged particle gyrating around the mean magnetic field). Overall, the stochastic character of the energization process points towards a Fermi-like acceleration mechanism (or DSA), driven by non-resonant interactions of the ion with the self-generated Bell's waves.\footnote{We remark that the detailed nature of the waves is not instrumental in the process of ion acceleration, since the Fermi mechanism only requires a sufficient  power in turbulent fluctuations, regardless of their origin. Yet, the details of wave-particle interactions (e.g., resonant versus non-resonant) may be important for the rate of acceleration.}

During the acceleration process, the ion gyrocenter moves predominantly along the mean magnetic field (in the $xz$ plane, for $\varphi=90\deg$), as shown by the time evolution of $x_i$ (blue) and $z_i$ (red) in panel (d). In addition, a clear drift can be seen in $y_i$ (green) towards negative values, i.e., in the direction of the background $\mathbf{E}_0$ (see SS09 for an analogous effect for positrons in pair shocks). 

As explained in \S\ref{sec:intro}, particles reflected by the shock can be accelerated by the background motional electric field $\mathbf{E}_0$ while drifting along the shock surface. Here, we do not differentiate between SDA and SSA, but we generically term this energization mechanism as ``direct acceleration''  (as opposed to ``diffusive acceleration'') or ``$\mathbf{E}_0$-driven acceleration''. We remind that $\mathbf{E}_0=-\mbox{\boldmath{$\beta$}}_0\times\mathbf{B}_0$ is the motional field at injection, and it does not include any contribution from self-generated turbulence.

\begin{figure}[tbp]
\begin{center}
\includegraphics[width=0.5\textwidth]{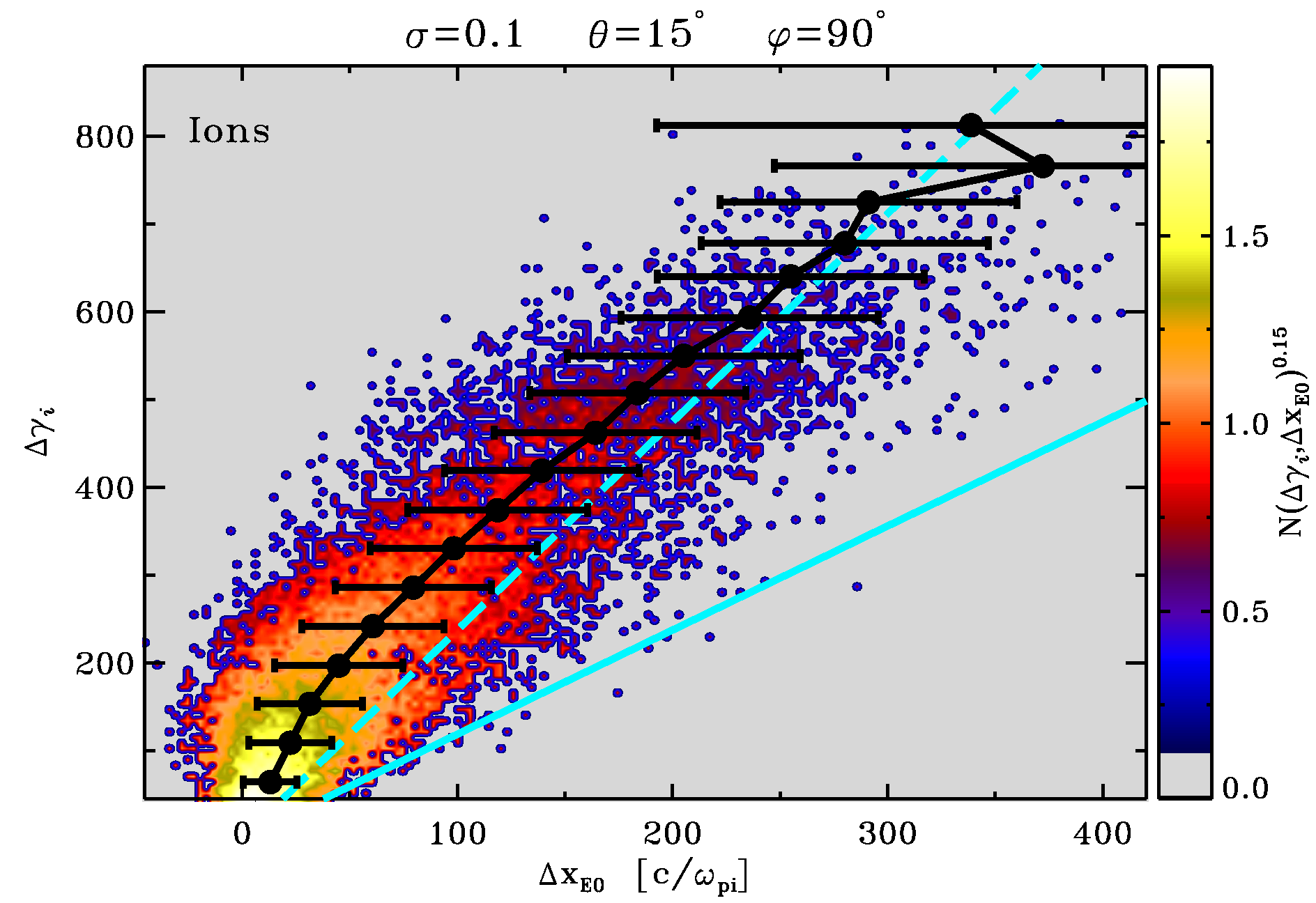}
\caption{Relative contribution of different acceleration mechanisms, for high-energy ions (such that $\gamma_i/\gamma_0>4$ at $\ompt=2250$) in a $\theta=15\deg$ subluminal shock. Here, $\Delta x_{\rm{E0}}$ is the ion displacement along $\mathbf{E}_0$, and $\Delta\gamma_i$ is the overall change in Lorentz factor. The 2D histogram shows the ion phase space density $N(\Delta\gamma_i,\Delta x_{\rm{E0}})$. For each bin in $\Delta\gamma_i$, the value of  $\Delta x_{\rm{E0}}$ averaged over the ion distribution is shown as a black dot with  $1\sigma$ error bars. The contribution $\Delta\gamma_{\rm E0}$ expected from direct acceleration by $\mathbf{E}_0$ is the solid cyan line. The dashed cyan line is the predicted $\Delta\gamma_{\rm E0}$ if the field $\mathbf{E}_0$ were twice as large.}
\label{fig:accel15}
\end{center}
\end{figure}

In \fig{accel15} we quantify the relative importance of this $\mathbf{E}_0$-driven acceleration mechanism with respect to the DSA process discussed above. For each of the high-energy ions extracted from the simulation, we measure the total energy gain $\Delta \gamma_i$ and the drift $\Delta x_{\rm{E0}}$ along the upstream field $\mathbf{E}_0$. For each bin in $\Delta \gamma_i$, the average value of $\Delta x_{\rm{E0}}$  is shown as a black dot with error bars. From $\Delta x_{\rm{E0}}$, we can compute the expected energy gain $\Delta\gamma_{\rm E0}=(q/mc^2)\,{E}_0\,\Delta x_{\rm{E0}}$ due to direct acceleration by the $\mathbf{E}_0$ field.\footnote{This formula assumes that most of the energy gain occurs in the upstream medium (where the background field is $\mathbf{E}_0$), which is correct since in the simulation frame no motional electric fields should persist in the downstream region.} We see that the resulting $\Delta\gamma_{\rm E0}$ (solid cyan line) is much lower than the actual energy gain (black dots), which suggests that DSA in Bell's waves plays a major role in the energization of ions. In fact, $\mathbf{E}_0$-driven acceleration contributes at most $\sim50\%$ of the total energy gain (see the dashed cyan line, which is the predicted $\Delta\gamma_{\rm E0}$ if the field $\mathbf{E}_0$ were twice as large).

\begin{figure}[tbp]
\begin{center}
\includegraphics[width=0.5\textwidth]{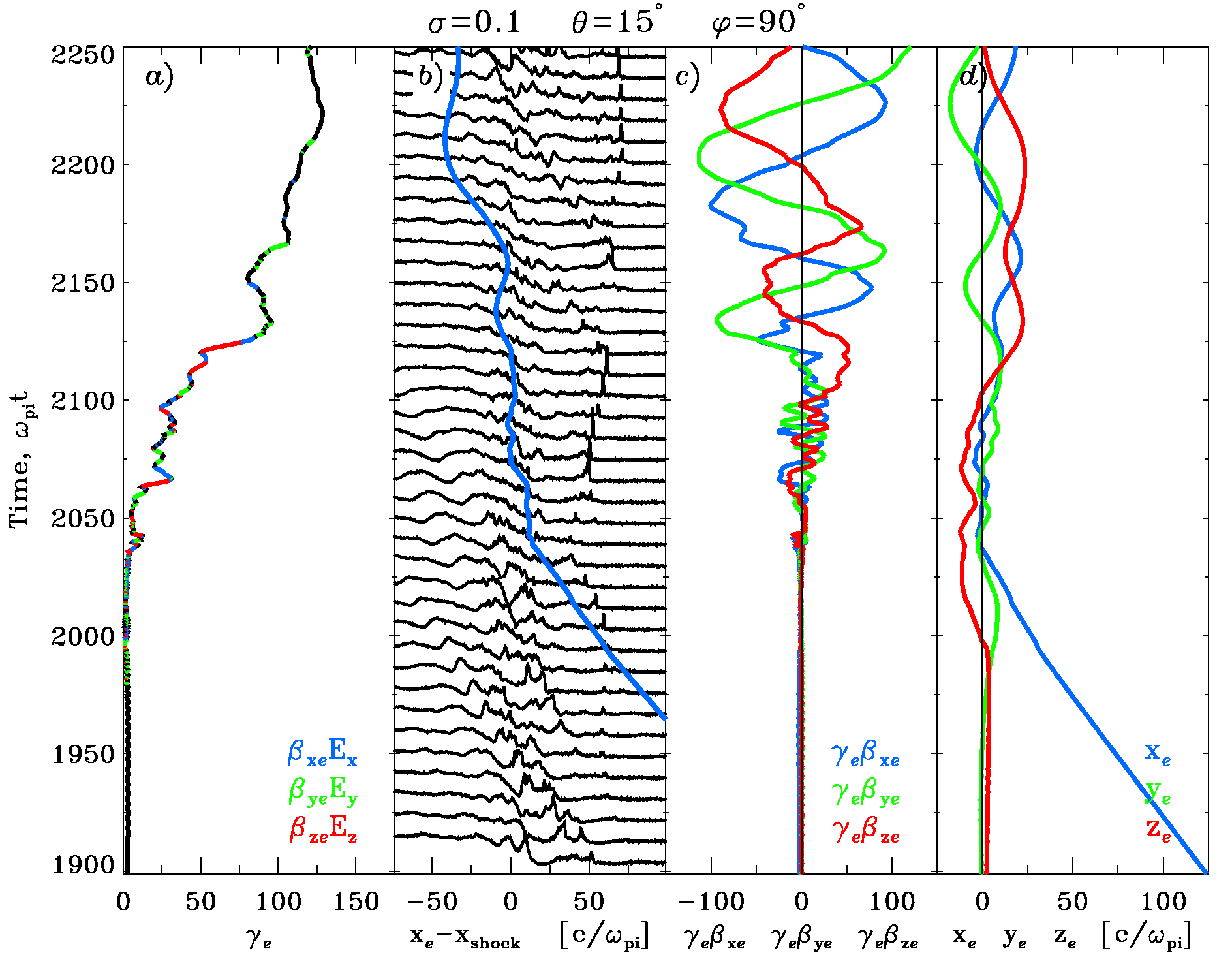}
\caption{Trajectory of a representative high-energy electron extracted from the simulation of a $\theta=15\deg$ subluminal shock. Panels as in \fig{ions15}, but here the fluid quantity plotted in panel (b) is the particle number density. Also, the electron Lorentz factor and dimensionless momentum have been divided by the mass ratio, although not indicated in the plot labels.}
\label{fig:lecs15}
\end{center}
\end{figure}

A separate argument needs to be made for electrons, whose acceleration path cannot be exactly the same as for ions, given the different spectrum (\fig{spectime15}), and above all the absence of returning electrons (compare phase spaces for ions and electrons in panels (j) and (k) of \fig{fluid15b}, respectively). In \fig{lecs15}, we plot the orbit of a representative high-energy electron in a $\theta=15\deg$ subluminal shock. The upstream trajectory of the electron gets perturbed (see panels (b) and (d) at $\ompt\sim2000$) even before entering the shock, by the density spike formed $\sim50\comp$ ahead of the shock by the pressure of returning ions. The electron energy (panel (a)) does not appreciably change before reaching the shock at $\ompt\sim2050$. Hereafter, the electron bounces several times across the shock transition layer (from $\ompt\sim2050$ to $\ompt\sim2175$), gaining energy by scattering off the Bell's waves (whose wavelength imprints the short-scale modulation seen in the 4-velocities of panel (c) for $2050\lesssim\ompt\lesssim2125$), similarly to what happens for ions.

The final energy of the selected electron ($\gamma_e m_e/m_i\sim120$) is much smaller than the downstream energy of the representative ion followed in \fig{ions15} ($\gamma_i\sim600$). This reflects the lower bulk energy (smaller by a factor of $\sim5$, see \fig{fluid15a}(d)) with which electrons enter the shock, with respect to ions. Due to their smaller Larmor radius, they are confined closer to the shock, and they do not penetrate very far upstream, where the electric field of Bell's waves would be stronger (or, in the standard Fermi picture, they do not sample the full velocity difference of the converging flows). Also, since they are more strongly tied to the magnetic field, they leave the acceleration region earlier by advection into the downstream (the electron in \fig{lecs15} stays at the shock for $\omega_{\rm pi}\Delta t\sim125$, as opposed to $\omega_{\rm pi}\Delta t\sim900$ for the ion in \fig{ions15}), which justifies the scarcity of returning electrons.\footnote{Electrons are also attracted into the downstream by the cross-shock electric field $E_{\rm x}>0$, created by electron-ion charge separation at the shock. We clearly see this effect for some  electrons, but not for the particle in \fig{lecs15}.} In summary, electrons display a lower energy gain per acceleration cycle (relative to ions), and a higher escape probability from the acceleration region. The combination of these two effects explains why in \fig{spectime15} electrons present a power-law tail much steeper than ions. 

As a side note, we point out that the density profiles stacked in panel (b) of \fig{lecs15} clearly show the process of shock reformation discussed in \S\ref{sec:fsub}. We see a density spike emerging ahead of the shock and then being advected into the shock on a typical timescale of $\sim100\,\omega_{\rm pi}^{-1}$. Then, a new spike forms, at a slightly larger distance (in fact, the shock transition region widens with time), and the whole process starts again.

%%%%%%%%%%%%%%%%%%%%%%%%%%%%%%%%%%%%%%%%
\subsection{Superluminal Shocks: $\thetacrit<\theta\leq90\deg$}\label{sec:asuper}

\subsubsection{Time Evolution}\label{sec:asuper_time}
\begin{figure}[tbp]
\begin{center}
\includegraphics[width=0.5\textwidth]{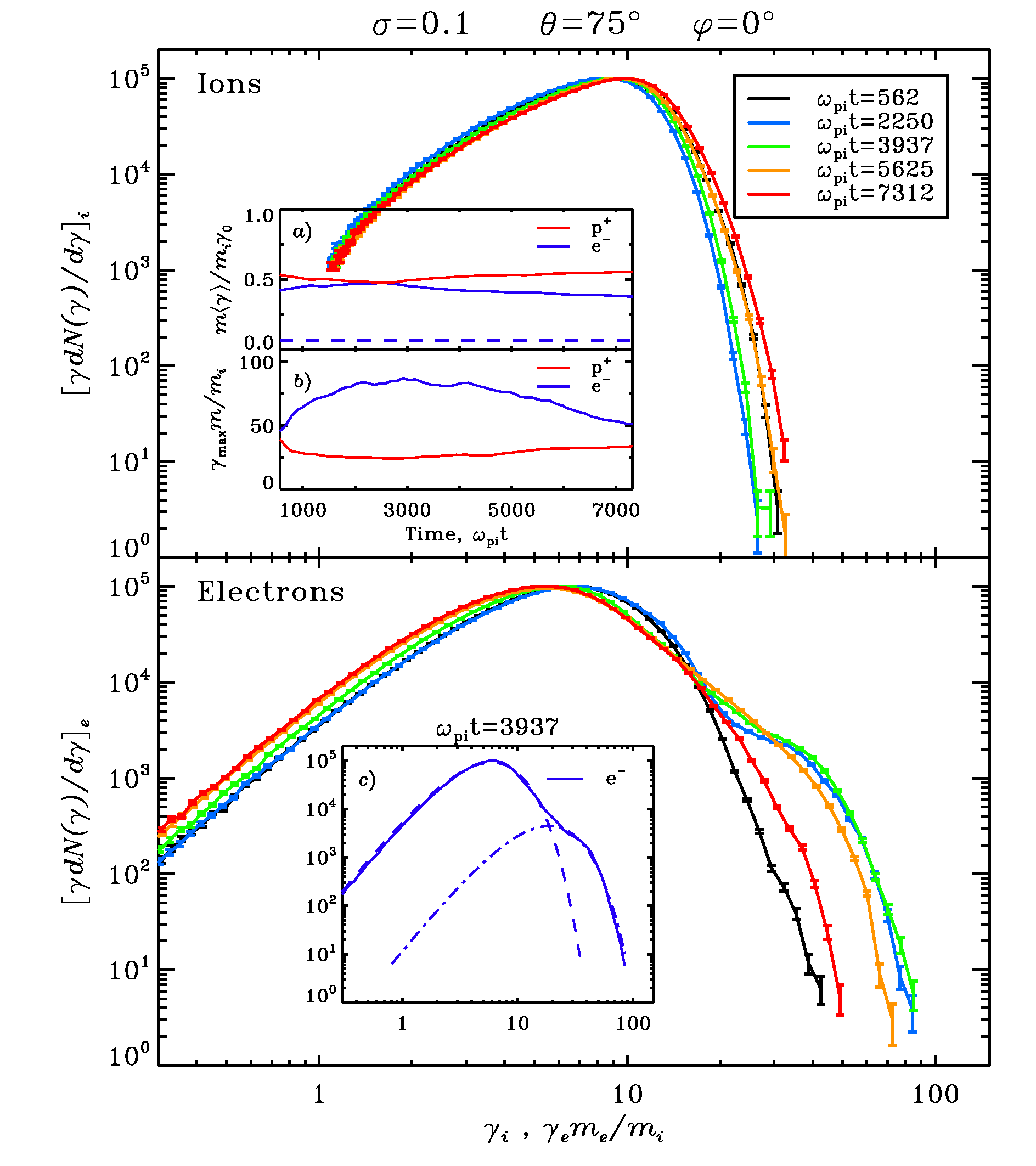}
\caption{Time evolution of the downstream energy spectrum in a $\theta=75\deg$ superluminal shock, for ions (upper panel) and electrons (lower panel): $\ompt=562$ (black), $\ompt=2250$ (blue), $\ompt=3937$ (green), $\ompt=5625$ (yellow), and $\ompt=7312$ (red). Subpanels as in the caption of \fig{spectime15}, but here the electron spectrum in panel (c) is fitted with two Maxwellians.}
\label{fig:spectime75}
\end{center}
\end{figure}

\fig{spectime75} follows the time evolution of the ion (upper panel) and electron (lower panel) downstream spectrum in a $\theta=75\deg$ superluminal shock. At all times, the ion spectrum is consistent with a purely Maxwellian distribution, i.e., nonthermal acceleration of ions is suppressed in superluminal shocks. The location of the thermal peak varies depending on the amount of ion energy transferred to electrons before the shock, via the mechanism described in \S\ref{sec:fsuper}.

The electron spectrum at early times ($\ompt=562$, black curve) is also compatible with a single Maxwellian distribution, but at later times (blue for $\ompt=2250$, green for $\ompt=3937$), a separate component emerges at high energies. We interpret this component as the result of extra \tit{heating} of some electrons by the wakefield oscillations described in \S\ref{sec:fsuper}, rather than as nonthermal \tit{acceleration}, for two main reasons. First, in any model of electron acceleration, we would expect a steady increase in the maximum electron Lorentz factor, at odds with what we observe in the blue line of panel (b). Also, we find that the downstream electron spectrum can be well fitted with two Maxwellians (panel (c)), with the hotter Maxwellian (dot-dashed line, with a temperature $\sim3$ times higher than the low-energy Maxwellian) accounting for the high-energy component. A fit that employs a low-energy Maxwellian plus a power law with exponential cutoff is less satisfactory.

The appearance of the high-energy component is accompanied by an increase in the average electron energy, at the expense of ions (panel (a) for $\ompt\lesssim3000$; red for ions, blue for electrons). For $\ompt\gtrsim3000$, the mean electron energy decreases (panel (a)), and the high-energy component recedes ($\ompt=5625$, yellow curve), until it finally disappears ($\ompt=7312$, red curve). The similarity between the time evolution of the average electron energy in panel (a) and the maximum electron Lorentz factor in panel (b) (a good proxy for the importance of the high-energy component) suggests that a common mechanism should regulate both the low-energy ``thermal'' part and the high-energy tail, as we now describe. 

At the beginning, electrons enter the shock with $x$-momentum $=-\gamma_0\beta_0\simeq-15$ and a small thermal spread. The synchrotron maser instability at the shock front produces powerful electromagnetic waves (Poynting flux in excess of $c\,E_0^2/4\pi$) that illuminate the upstream flow, causing efficient transfer of bulk momentum from ions to electrons (see \S\ref{sec:fsuper}). As electrons with larger bulk energy enter the shock, the synchrotron maser instability is more efficient in producing strong precursor waves, and the transfer of energy from ions to electrons in the upstream becomes even more significant. This self-reinforcing cycle repeats until electrons reach energy equipartition with ions (panel (a), at $\ompt\sim2500$). As the strength of the electromagnetic precursor increases, wakefield oscillations of larger amplitude are generated in the upstream \citep{lyubarsky_06}. As anticipated in \S\ref{sec:fsuper}, their dissipation then results in a more pronounced high-energy component in the electron spectrum. This explains why the maximum electron Lorentz factor increases for $\ompt\lesssim3000$ (blue line in panel (b)).

Heated by the upstream wakefield oscillations, electrons now enter the shock much hotter than before, i.e., their comoving momentum spread is larger. In such conditions, the growth of the synchrotron maser instability is quenched \citep{hoshino_91}, and the electromagnetic precursor becomes less powerful. This reduces the transfer of bulk energy from ions to electrons in the upstream, causing the decrease in average electron energy observed at $\ompt\gtrsim3000$ in panel (a).\footnote{This effect has been verified with controlled numerical experiments, in which we initialized the incoming particles with a larger thermal spread than the usual $\Delta\gamma=10^{-4}$. For $\Delta\gamma\gtrsim\pow{1}$, both the power of the electromagnetic precursor and the upstream ion-to-electron energy transfer are gradually suppressed, as well as the high-energy component in the electron spectrum.} Also, the upstream wakefield oscillations now become weaker, so that fewer electrons are injected into the high-energy spectral component, and the maximum electron Lorentz factor decreases (panel (b), at $\ompt\gtrsim3000$). 

In our simulation, we find that the electron spectrum approaches a steady state for $\ompt\gtrsim7000$ (we followed the shock evolution up to $\ompt\sim10,000$). However, since the incoming electrons are now colder, we cannot exclude that at later times the whole process may repeat, with efficient transfer of energy to electrons accompanied by the emergence of a new high-energy spectral component. Instead, we do not expect the ion spectrum to deviate from a Maxwellian distribution, since ions are not significantly affected by the upstream wakefield oscillations, due to their larger inertia. 

Finally, we comment on the difference between our 2D simulations and the 1D results presented by \citet{hoshino_08}, that reported nonthermal \tit{acceleration} of electrons in perpendicular magnetized electron-ion shocks. First, we find that 1D simulations over-estimate the power of the electromagnetic precursor, since they cannot resolve variations along the shock surface (\fig{fluid75b}(c) shows that the strength of the electromagnetic precursor is not uniform in $y$). As a result, 1D runs tend to over-estimate the efficiency of ion-to-electron energy transfer, and the importance of the high-energy component in the electron spectrum. For $m_i/m_e=16$, computational domains with more than $64$ transverse cells are required to obtain consistent 2D results.\footnote{In fact, to follow the spectral evolution shown in \fig{spectime75}, we have used a simulation box with 64 transverse cells, which is small enough to suppress the growth of the numerical instability mentioned in \S\ref{sec:setup}, yet large enough to preserve the relevant 2D properties of the shock.} Second, \citet{hoshino_08} followed the shock evolution only up to $\ompt\sim850$, a timespan too short to appreciate the saturation and decrease in the maximum electron Lorentz factor shown in panel (b). As mentioned above, this piece of evidence is essential for our interpretation of the high-energy spectral component  as a result of wakefield \tit{heating}, rather than \tit{acceleration}.

\subsubsection{Energization Mechanism}\label{sec:asuper_mech}
\begin{figure}[tbp]
\begin{center}
\includegraphics[width=0.5\textwidth]{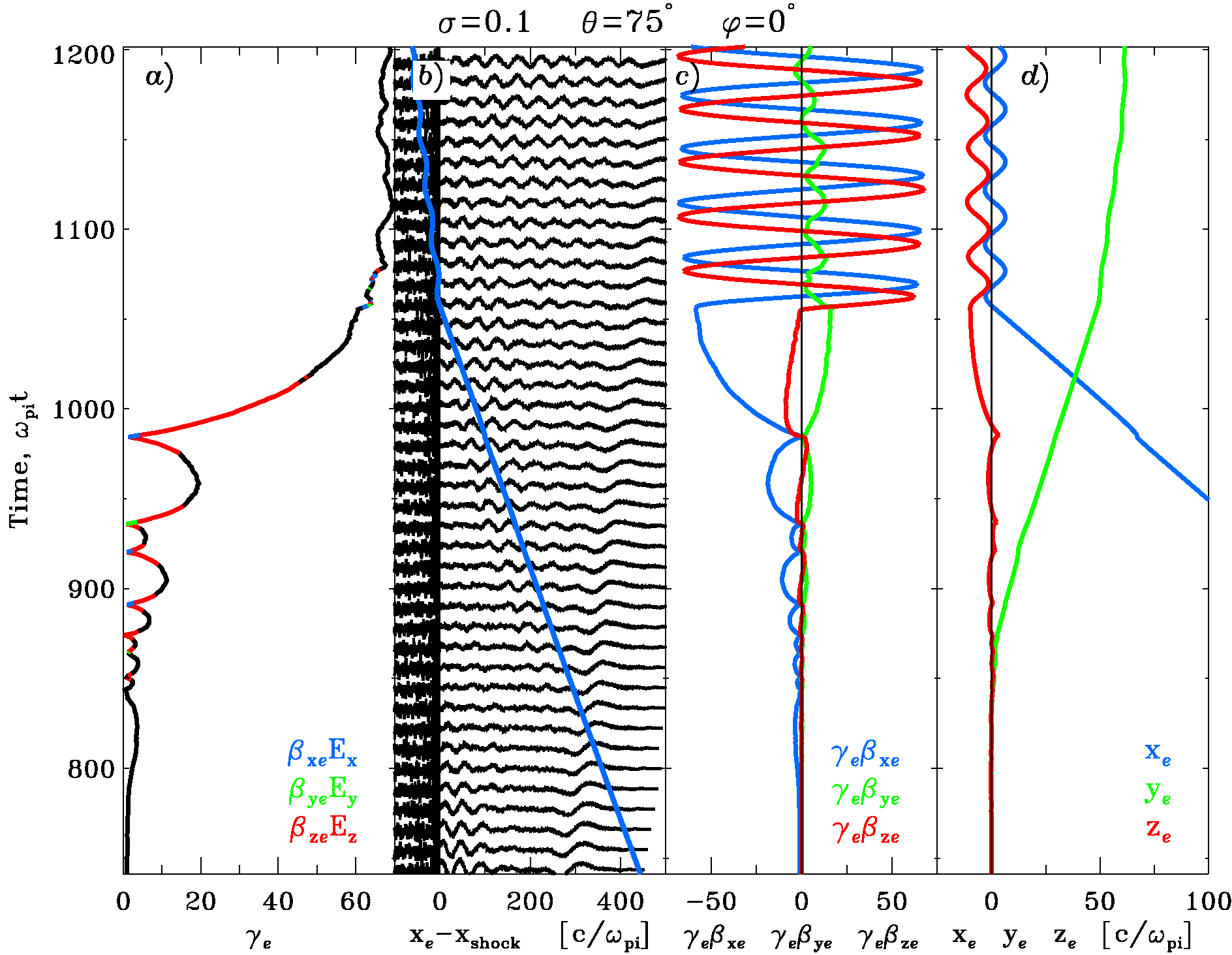}
\caption{Trajectory of a representative high-energy electron extracted from the simulation of a $\theta=75\deg$ superluminal shock. Panels as in \fig{ions15}, but here the fluid quantity plotted in panel (b) is $E_{\rm x}$. Also, the electron Lorentz factor and dimensionless momentum have been divided by the mass ratio, although not indicated in the plot labels. We remind that in this case ($\varphi=0\deg$), the upstream background electric field $\mathbf{E}_0$ is along $+\mathbf{\hat{z}}$.}
\label{fig:lecs75}
\end{center}
\end{figure}

The trajectory of a representative high-energy electron in a superluminal shock ($\theta=75\deg$)  is presented in \fig{lecs75}. Panel (a) shows that all of the energy gain, up to $\gamma_em_e/m_i\sim70$, occurs in the upstream region \tit{before} the electron encounters the shock at $\ompt\sim1050$ (see the electron $x$-location relative to the shock in panel (b)). After that, the electron is advected downstream, with no appreciable change in energy. 

For $\ompt\lesssim850$, the electron is first boosted and then deboosted (see energy, panel (a); or $x$-momentum, blue line in panel (c)), in response to the wakefield wiggle seen at $\sim350\comp$ ahead of the shock ($E_{\rm x}$ in panel (b)). So far, the trajectory of the electron in \fig{lecs75} presents no qualitative difference from  the bulk of incoming electrons. But at $\ompt\sim850$, presumably due to a strong upstream-oriented kick imparted by the wakefields, the energy of the selected electron becomes so small that it can easily decouple from the bulk flow. As anticipated in \S\ref{sec:fsuper}, this is the ``injection'' step for all the electrons that will end up in the high-energy spectral component seen in \fig{spectime75}.

After decoupling from the bulk flow, the electron gyrocenter starts moving with drift velocity $\mathbf{E}_0\cross\mathbf{B}_0/B_0^2$, which for the field configuration employed here ($\varphi=0\deg$) corresponds to a straight path in the $xy$ plane (see $y_e$ in panel (d) for $\ompt\gtrsim850$). At the same time, the electron gets pulled by the upstream motional electric field $\mathbf{E}_0$, and it performs a series of Larmor cycles (see energy in panel (a) and 4-velocity in panel (c), for $\ompt\gtrsim850$). When the electron velocity is aligned with $-\mathbf{E}_0$ the particle energy grows, but it decreases by the same amount when the two are oppositely directed (see red segments in panel (a)). Overall, no \tit{net} energy gain is expected from $\mathbf{E}_0$ during a \tit{complete} Larmor cycle. In fact, as seen from the upstream frame, the electron motion is just a Larmor gyration around the background magnetic field, with energy determined by the magnitude of the kick imparted by the wakefields.

The final electron energy is determined by the combination of two factors: (\tit{i}) the amplitude of the energy oscillations seen in panel (a); and (\tit{ii}) the gyro-phase with which the particle enters the shock. Regarding (\tit{i}), secular variations (i.e., averaged over gyro-phase) in the electron energy are determined by the kicks imparted by the wakefields. In particular, the electron will be most sensitive to such kicks when its energy is smaller. At this point, as seen from the upstream frame, the electron is moving away from the shock. If $E_{\rm x}<0$, the electron will receive a kick in the same direction of its momentum, and its upstream-frame energy will increase. As seen from the downstream frame, this corresponds to a larger amplitude of the resulting energy oscillation (e.g., at $x\sim960\comp$ in panel (a)). The opposite happens if the phase of wakefield oscillations is such that $E_{\rm x}>0$, so that the following energy oscillation will have a smaller amplitude (e.g., at $x\sim900\comp$ in panel (a)). As a confirmation of the importance of wakefields, we see that the amplitude of the energy oscillations in panel (a) increases dramatically for $\ompt\gtrsim925$, when the selected electron enters a region with strong wakefields ($E_{\rm x}$ in panel (b), following the particle track for $\ompt\gtrsim925$). 

On the other hand, depending on the gyro-phase with which the particle encounters the shock (point (\tit{ii}) above), the electron will end up in a different energy bin of the downstream high-energy Maxwellian (dot-dashed line in \fig{spectime75}(c)).

Finally, we remark that, despite the significant wakefield heating experienced in the upstream, which could possibly serve as pre-injection for a Fermi-like process, the electron in \fig{lecs75} does not bounce back from the shock into the upstream, due to insufficient turbulence on the downstream side. In fact, in our simulations of superluminal shocks, we do not find any evidence for returning electrons  (see the electron phase space in \fig{fluid75a}(f)), confirming that Fermi acceleration is suppressed for superluminal configurations.

%%%%%%%%%%%%%%%%%%%%%%%%%%%%%%%%%%%%%%%%%%%
\section{Dependence on Upstream Conditions}\label{sec:survey}
Armed with a better understanding of the energization mechanisms that operate in relativistic magnetized electron-ion shocks, we now explore how the efficiency of ion-to-electron energy transfer and of particle acceleration depends on the conditions of the upstream flow. First, keeping the bulk Lorentz factor ($\gamma_0=15$) and magnetization ($\sigma=0.1$) fixed, as we did so far, we investigate the transition between subluminal and superluminal shocks by exploring the full range of magnetic obliquities, from $\theta=0\deg$ to $\theta=90\deg$ (in \S\ref{sec:spectheta}). Then, for the two representative obliquities discussed above (one subluminal, $\theta=15\deg$; one superluminal, $\theta=75\deg$), we extend our analysis to different values of the magnetization (in \S\ref{sec:specsigma}) and bulk Lorentz factor (in \S\ref{sec:specgamma}).

The comparisons in this section are performed at $\ompt=2250$. Yet, as shown in Figs.~\fign{spectime15} and \fign{spectime75}, the downstream particle spectrum is not in steady state. In subluminal shocks (\fig{spectime15}), the nonthermal tail grows with time, so that the acceleration efficiency we compute at $\ompt=2250$  should be taken as a lower limit. In superluminal shocks (\fig{spectime75}), the electron high-energy component is most pronounced between $\ompt\sim2000$ and $\ompt\sim3000$, and then it disappears, although a cyclical behavior is likely, yet undemonstrated (see \S\ref{sec:asuper_time}). At $\ompt=2250$, we are capturing this high-energy component when it is close to its maximum, and more likely to result in an observational signature.

In the following, we fit the downstream particle spectrum with a three-dimensional Maxwellian plus a power-law tail with an exponential cutoff. This is not strictly appropriate for superluminal magnetized shocks, where the high-energy component of the electron spectrum is better fitted with a second (hotter) Maxwellian, as discussed in  \S\ref{sec:asuper_time}. Although this  component actually results from \tit{heating} (rather than \tit{acceleration}), in this section we treat it as a nonthermal tail, for easier comparison with observations.\footnote{Observationally, a suprathermal component in the radiation spectrum will most likely be ascribed to nonthermal acceleration of particles, and fitted as a power-law.}  

The form chosen for our fitting function allows to split the particle population into a \tit{thermal} and a \tit{nonthermal} component, and to follow how the relative importance of the two sub-populations changes with the upstream parameters. In Figs.~\fign{spectheta}-\fign{specgam75}, panel (a) presents the mean downstream particle energy (red for ions, blue for electrons) in units of the bulk energy of injected ions, including \tit{both} thermal \tit{and} nonthermal particles. Panels (b)-(d) provide a better characterization of the nonthermal component: panel (b) shows the slope $p$ of the best-fitting power law, such that $dN_{\msc{pl}}(\gamma)/d\gamma\propto \gamma^{-p}$; panels (c) and (d) present respectively the fraction of particles and energy contained in the nonthermal tail (red for ions and blue for electrons, relative to the total number or energy \tit{of that species}).\footnote{For instance, to obtain the fraction of energy in nonthermal \tit{electrons} relative to the bulk energy of injected \tit{ions} (which is often required in phenomenological models of GRBs and AGN jets), the value read from the blue line in panel (d) should be multiplied by the value of the blue line in panel (a).} In subpanels (b)-(d), dotted lines (instead of the usual solid lines) mark the cases in which a nonthermal power-law tail is \tit{not} a physically-motivated model for the high-energy component (i.e., for electrons in superluminal magnetized shocks, as discussed above).

%%%%%%%%%%%%%%%%%%%%%%%%%%%%%%%%%%%%%%%%

\subsection{Dependence on Magnetic Obliquity $\theta$}\label{sec:spectheta}

\begin{figure}[tbp]
\begin{center}
\includegraphics[width=0.5\textwidth]{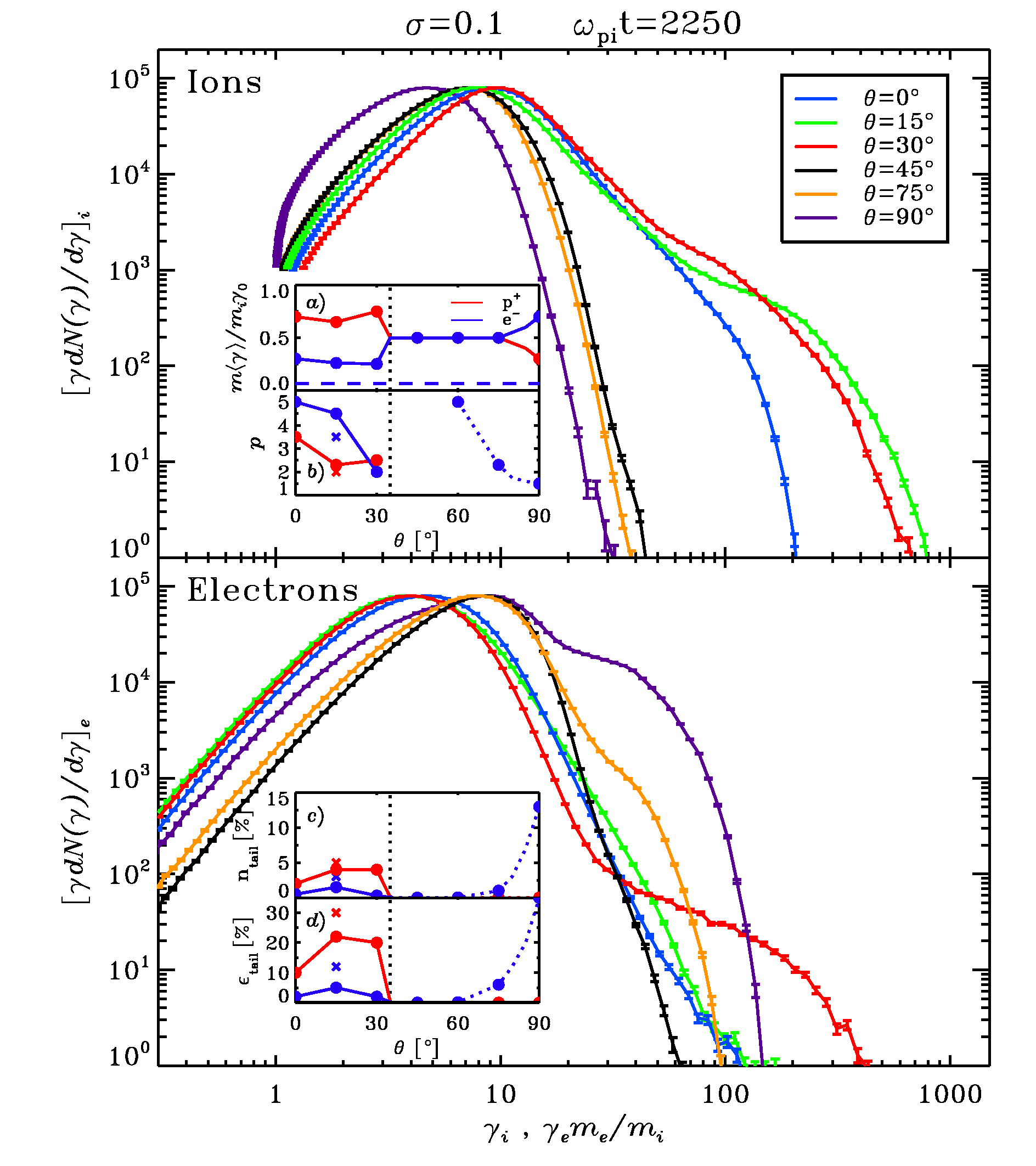}
\caption{Downstream particle spectrum at $\ompt=2250$ (upper panel for ions, lower panel for electrons) for different magnetic obliquities, from $\theta=0\deg$ (parallel shock) to $\theta=90\deg$ (perpendicular shock). In subpanels, subluminal and superluminal geometries are separated by a vertical dotted black line, corresponding to the critical angle $\thetacrit\simeq34\deg$. Subpanels: (a) downstream ion (red) and electron (blue) average energy, in units of the bulk energy of injected ions, with the horizontal dashed blue line showing the value expected for electrons in the absence of any ion-to-electron energy transfer ($=m_e/m_i\simeq0.06$); (b) power-law slope of the nonthermal tail (red for ions, blue for electrons); (c) fraction of ions (red) or electrons (blue) in the nonthermal tail; (d) fraction of energy in the ion (red) or electron (blue) nonthermal tail, with respect to the total kinetic energy of that species. In subpanels (b)-(d), dotted blue lines  mark the cases in which a nonthermal power-law tail is not a physically-motivated model for the electron high-energy component (see text). Crosses for $\theta=15\deg$ (red for ions, blue for electrons) correspond to values measured at $\ompt=5062$ (instead of $\ompt=2250$), to show how the nonthermal tail changes in time.
}
\label{fig:spectheta}
\end{center}
\end{figure}

\fig{spectheta} presents the downstream particle spectrum at $\ompt=2250$ (upper panel for ions, lower panel for electrons), for a sample of magnetic obliquities covering the full range from $\theta=0\deg$ to $90\deg$. In this section, we describe how the efficiency of particle acceleration and electron heating depends on magnetic inclination, emphasizing the importance of the critical boundary $\thetacrit\simeq34\deg$ between subluminal and superluminal configurations. 

The ion spectrum in subluminal shocks (blue curve for $\theta=0\deg$, green for $\theta=15\deg$, red for $\theta=30\deg$) shows a prominent nonthermal tail at high energies, together with a thermal bump at low energies. Instead, for superluminal shocks (black for $\theta=45\deg$, yellow for $\theta=75\deg$, purple for  $\theta=90\deg$), the ion spectrum does not appreciably deviate from a Maxwellian distribution. As shown in panels (c) and (d), the efficiency of ion acceleration (red lines) abruptly drops between $\theta=30\deg$ and $\theta=35\deg$, as the magnetic obliquity passes the critical value $\thetacrit\simeq34\deg$. 

For subluminal configurations, the ion nonthermal tail shows a non-monotonic behavior with angle. Ion acceleration is most efficient for $\theta=15\deg$, with a fraction $\sim4\%$ of particles and $\sim22\%$ of energy contained in the tail at $\ompt=2250$ (red dots for $\theta=15\deg$ in panels (c) and (d), respectively). As discussed in \S\ref{sec:asub_time}, at $\ompt=5062$ the ion acceleration efficiency reaches as much as $\sim5\%$ by number and $\sim30\%$ by energy (red crosses in panels (c) and (d)). By comparison, at $\ompt=2250$ the ion nonthermal tail for $\theta=0\deg$ only accounts for $\sim2\%$ of particles and $\sim10\%$ of ion energy, whereas the acceleration efficiency for $\theta=30\deg$ ($\sim4\%$ by number, $\sim20\%$ by energy) is comparable to $\theta=15\deg$, but the tail extends to smaller Lorentz factors. Also, the ion tail is flatter for $\theta=15\deg$, with a slope $p\sim2.2$, compared to $p\sim3.5$ for $\theta=0\deg$ and $p\sim2.5$ for $\theta=30\deg$ (red line in panel (b)).

At earlier times (e.g., $\ompt=1125$, not shown here), the trend was different, with larger angles (yet still subluminal) showing flatter ion tails that extended to higher Lorentz factors and contained more particles and energy. In fact, this was the hierarchy that SS09 observed in electron-positron shocks (see Fig.~11 in SS09, same colors as here). In SS09, such trend was attributed to a change in the dominant acceleration mechanism, with DSA prevailing in quasi-parallel shocks and SDA most efficient for $15\deg\lesssim\theta\lesssim\thetacrit$. In electron-ion shocks, as we have shown in \S\ref{sec:asub_mech}, magnetic turbulence created by the returning ions (in the form of Bell's waves) can dramatically affect the acceleration process, much more than in electron-positron shocks, where the amplitude of self-generated upstream turbulence was much lower (see SS09). The hierarchy seen in the subluminal ion spectra of \fig{spectheta} will then reflect the influence of magnetic inclination on the strength of Bell's waves. 

Since the gyrocenters of returning particles tend to move along the upstream field, for higher (yet subluminal) obliquities the returning ions will be confined closer to the shock. This has two opposite consequences. First, since the returning ions are responsible for triggering Bell's instability,  at larger obliquities its growth is more likely to be  suppressed by advection into the shock. In fact, at $\ompt=2250$, Bell's modes are already fully developed for $\theta=15\deg$, whereas for $\theta=30\deg$ they are just starting to grow, and the acceleration process is still governed by SDA, as in electron-positron flows (SS09). On the other hand, Bell's instability saturates when the self-generated magnetic energy approaches the energy density of returning ions \citep{riquelme_09}, which is larger for higher obliquities since the returning particles accumulate in a smaller region ahead of the shock, as discussed above. It follows that Bell's waves are stronger for $\theta=15\deg$ than for $\theta=0\deg$. Overall, the best compromise at $\ompt=2250$ is achieved for $\theta=15\deg$, where the waves are both powerful and rapidly-growing (compared to advection).

At later times, the situation may change. The growth of Bell's waves for $\theta=30\deg$ will not be limited by advection any longer, and the turbulence will saturate at higher amplitudes than for $\theta=15\deg$, giving efficient ion acceleration via DSA. Therefore, the trend in acceleration efficiency observed for subluminal electron-positron shocks by SS09 may eventually be restored for ions in electron-ion shocks as well. However, here it will be driven by the increase with angle in the self-generated turbulence that mediates DSA, whereas in electron-positron flows it follows the amplitude of the background motional electric field responsible for SDA.

As discussed in \S\ref{sec:asub}, acceleration of electrons in subluminal shocks is depressed with respect to ions. For nearly-parallel shocks (blue curve for $\theta=0\deg$, green for $\theta=15\deg$), a minor fraction ($\sim1\%$) of the incoming electrons are accelerated during their passage through the shock, as explained in \S\ref{sec:asub_mech}. In the downstream, they populate a steep ($p\sim5$) power-law tail which only contains $\sim3\%$ of the total electron energy. The tail may get somewhat flatter with time, thus accounting for more particles and energy (see blue crosses in panels (b)-(d), for $\theta=15\deg$), but the acceleration efficiency for electrons remains smaller than for ions by at least a factor of three. Close to the critical obliquity $\thetacrit\simeq34\deg$ (see $\theta=30\deg$, red curve), a more pronounced tail appears in the electron spectrum, populated by particles accelerated at the shock via the SDA mechanism, in analogy to what happens in electron-positron flows (SS09). However, despite extending to high energies (up to $\gamma_e m_e/m_i\sim400$ at $\ompt=2250$), the electron tail for $\theta=30\deg$ only contains $\sim1\%$ of particles and $\sim2\%$ of electron energy. 

Instead, a prominent high-energy tail is observed in the electron spectrum of superluminal shocks, beyond the thermal bump. As the obliquity angle increases from $\theta=75\deg$ (yellow curve) to $\theta=90\deg$ (purple), the tail stretches to higher Lorentz factors (maximum $\gamma_e m_e/m_i $ from $\sim95$ to $\sim150$), and it contributes a larger fraction of electrons (from $\sim2\%$  to $\sim13\%$) and electron energy (from $\sim6\%$ to as much as $\sim35\%$). If this high-energy component is fitted with a power law (but a fit with a second Maxwellian is actually more physically motivated, see \S\ref{sec:asuper_time}), we find that it gets flatter for higher obliquities (from $p\sim2.3$ for $\theta=75\deg$ to $p\sim1.5$ for $\theta=90\deg$).

As explained in \S\ref{sec:asuper_mech}, this high-energy component results from extra heating of some electrons by the upstream wakefield oscillations described in \S\ref{sec:fsuper}, that are ultimately generated by the radiative push of the precursor wave on the incoming flow. Since the amplitude of wakefield oscillations increases with the strength of the precursor wave \citep{lyubarsky_06}, which is larger for higher obliquities (roughly $\propto E_0^2\propto\sin^2\theta$), the electron high-energy component will be more pronounced for quasi-perpendicular shocks, as seen in \fig{spectheta}. Being due to \tit{heating}, rather than \tit{acceleration}, we do not expect the electron tail in superluminal shocks to extend with time to higher Lorentz factors. Rather, it may recede to lower energies, and possibly be hidden by the low-energy thermal bump, as shown in \S\ref{sec:asuper_time} for $\theta=75\deg$.

In superluminal shocks, the electromagnetic precursor wave also causes efficient transfer of  energy from ions to electrons ahead of the shock (see \S\ref{sec:fsuper}). As a result, in the downstream region the electron and ion average energies are comparable (see \fig{spectheta}(a), for $35\deg\lesssim\theta\lesssim75\deg$), or electrons may be even hotter than ions (for $75\deg\lesssim\theta\lesssim90\deg$).  In subluminal shocks ($0\deg\lesssim\theta\lesssim30\deg$), electrons are boosted and heated ahead of the shock, as a result of the perturbation induced on the incoming flow by the cloud of returning ions (see  \S\ref{sec:fsub}). With respect to superluminal angles, here the downstream electron energy is a smaller fraction ($\sim20\%-30\%$) of the initial ion bulk energy, but still much in excess of $m_e/m_i\simeq6\%$ (for $m_i/m_e=16$), the value we would expect in the absence of any ion-to-electron energy transfer. 

In Appendix \ref{sec:specmime}, we show that the values quoted here for the efficiency of electron heating and of particle acceleration are essentially unchanged for larger (and more realistic) mass ratios.

%\vspace{-0.3in}
%%%%%%%%%%%%%%%%%%%%%%%%%%%%%%%%%%%
\subsection{Dependence on Magnetization $\sigma$}\label{sec:specsigma}
In this section, we explore the dependence of shock thermalization and acceleration upon the magnetization of the upstream flow. We keep the magnetic obliquity fixed and vary the magnetization  from $\sigma=10^{-5}$, a virtually unmagnetized shock, up to $\sigma=1.0$, where the upstream field is in equipartition with the kinetic energy of injected particles. We discuss separately the cases of subluminal (in \S\ref{sec:msub}, for $\theta=15\deg$) and superluminal (in \S\ref{sec:msuper}, for $\theta=75\deg$) shocks.

\begin{figure}[tbp]
\begin{center}
\includegraphics[width=0.5\textwidth]{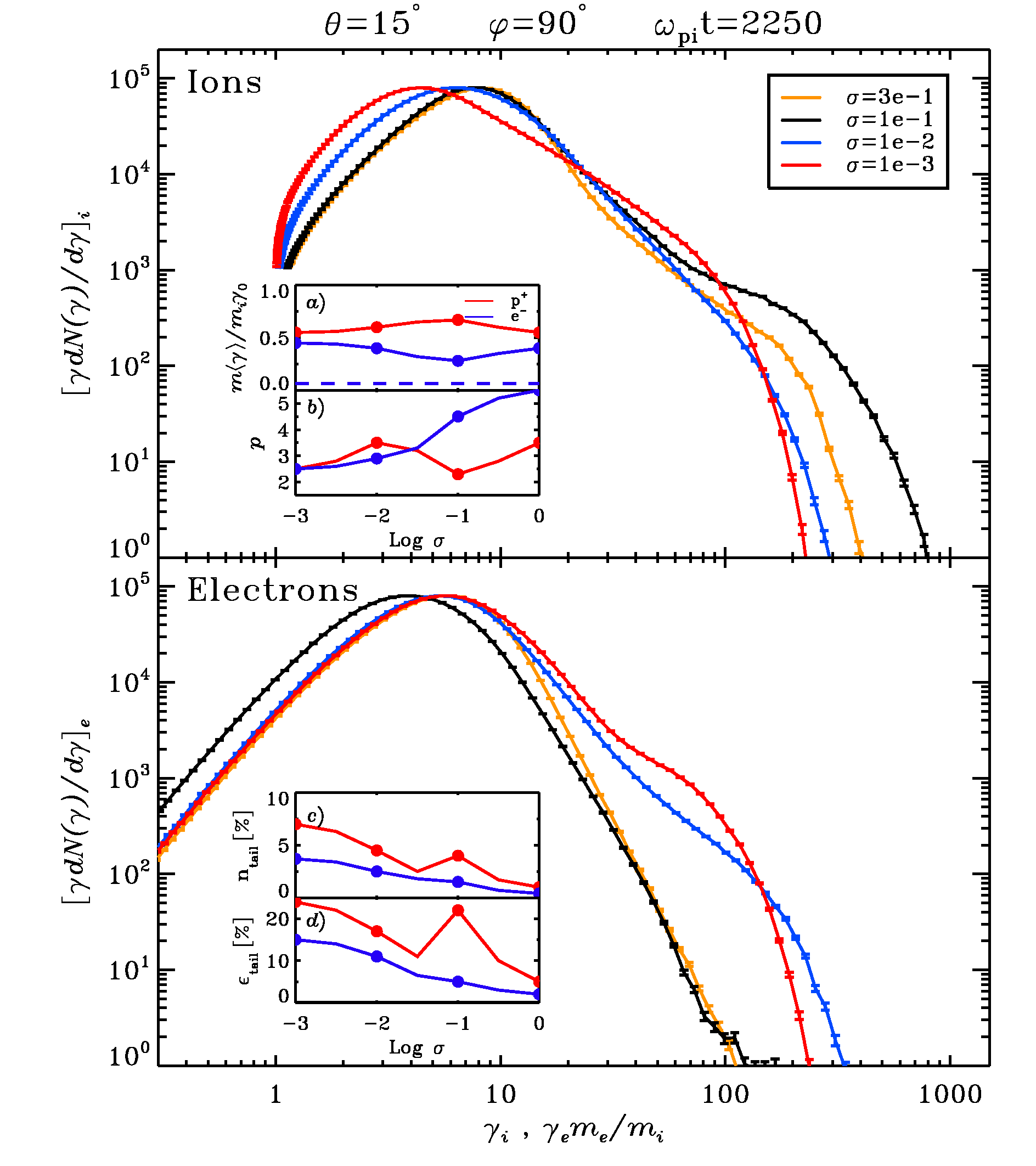}
\caption{Downstream particle spectrum at $\ompt=2250$ (upper panel for ions, lower panel for electrons) for different magnetizations. We fix the magnetic obliquity $\theta=15\deg$ (subluminal shock). Subpanels as in \fig{spectheta}.}
\label{fig:specsig15}
\end{center}
\end{figure}

\subsubsection{Subluminal Shocks: $0\deg \leq\theta<\thetacrit$}\label{sec:msub}
\fig{specsig15} presents the downstream particle spectrum of a $\theta=15\deg$ subluminal shock, for different magnetizations. The ion spectrum (upper panel) shows in all cases a pronounced nonthermal tail, whereas a significant high-energy component appears for electrons (lower panel) only at $\sigma\lesssim10^{-2}$. This corresponds to a change in the mechanism that mediates the shock. 

For $\sigma\lesssim10^{-2}$ (blue curve for $\sigma=\pow{2}$, red for $\sigma=\pow{3}$), the shock is mediated by the filamentation (Weibel) instability, as in strictly unmagnetized flows \citep{spitkovsky_05,spitkovsky_08}. The free energy for the instability comes from the counter-streaming between the incoming plasma and the shock-accelerated particles that propagate ahead of the shock. In turn, magnetic turbulence generated by the instability mediates nonthermal acceleration of particles at the shock front, via a Fermi-like mechanism \citep[][]{spitkovsky_08b,martins_09}. 

In fact, the ion spectrum in such shocks shows a prominent nonthermal tail. As the magnetization decreases below $\sigma=\pow{2}$, the tail becomes flatter (from $p\sim3.5$ at $\sigma=10^{-2}$ to $p\sim2.5$ at $\sigma=10^{-3}$, red line in \fig{specsig15}(b)) and it contains a larger fraction of ions (from $\sim5\%$ to $\sim7\%$, \fig{specsig15}(c)) and ion energy (from $\sim17\%$ to $\sim24\%$, \fig{specsig15}(d)). Similarly, the electron spectrum for $\sigma\lesssim\pow{2}$ presents a significant high-energy tail, comparable to the ion tail in slope (compare red and blue lines in panel (b), for $\sigma\lesssim\pow{2}$) and maximum Lorentz factor. The acceleration efficiency for electrons is not much smaller than for ions, at most a factor of two (compare red and blue lines in panels (c) and (d), for $\sigma\lesssim\pow{2}$).\footnote{For both electrons and ions, the downstream spectrum at even lower magnetizations (we explored down to $\sigma=10^{-7}$) is identical to the case $\sigma=10^{-3}$ shown here, which is therefore a good representation of weakly magnetized (and unmagnetized) shocks.}

The similarity between the acceleration properties of electrons and ions in weakly magnetized flows ($\sigma\lesssim\pow{2}$) comes from the fact that the two species enter the shock with comparable bulk energy (compare red and blue lines in panel (a), for $\sigma\lesssim\pow{2}$), due to efficient transfer of energy from ions to electrons ahead of the shock (Spitkovsky et al., in prep). The Fermi mechanism, which only depends on the particle  Larmor radius (i.e., on particle energy, rather than mass), will then proceed in a similar way for ions and electrons.

The situation changes for $\sigma\gtrsim\pow{2}$. Here, the magnetic field is strong enough so that the upstream plasma cannot filament, and the growth of Weibel instability is suppressed. In these conditions, the current of shock-accelerated ions moving ahead of the shock can trigger Bell's instability, as discussed in \S\ref{sec:fsub}. For $\pow{2}\lesssim\sigma\lesssim0.3$, the waves generated by Bell's mechanism are responsible for mediating the shock and driving the acceleration of ions. They are most powerful for $\sigma\sim0.1$ (black curve), which then corresponds to a peak in the ion acceleration efficiency ($\sim4\%$ by number, $\sim22\%$ by energy) and a flatter nonthermal tail ($p\sim2.2$). 

In contrast, acceleration of electrons is depressed in highly magnetized flows (black curve for $\sigma=0.1$), as discussed in \S\ref{sec:asub}. As the magnetization increases, returning particles are confined closer to the shock (in units of $\comp$, but same distance in Larmor radii), so their perturbation to the incoming flow extends over a smaller length. It follows that the transfer of energy from ions to electrons in the upstream has less time to proceed, and it is therefore less efficient (see panel (a), from $\sigma=\pow{3}$ to $\sigma=0.1$). For $\sigma=0.1$, electrons enter the shock with $\sim15\%$  of the initial ion energy (see \S\ref{sec:fsub}), so their Larmor radius is small compared to ions.  Then, only a small fraction of electrons ($\sim1\%$, panel (c)) are accelerated at the shock before being advected downstream, where they populate a steep nonthermal tail ($p\sim4.5$, panel (b)).

At even higher magnetizations ($\sigma\gtrsim0.3$, yellow curve for $\sigma=0.3$), the instability that mediates the shock changes again, and the dominant mode is now generated via gyro-frequency resonance with the returning ions (see Appendix \ref{sec:comparison}). However, the overall shape of the electron and ion spectrum does not significantly change (compare black, for $\sigma=0.1$, and yellow, for $\sigma=0.3$), showing that, despite the different polarization of the dominant mode (now resonant, as opposed to Bell's non-resonant waves), the physics of the acceleration is not significantly altered.  However, the level of magnetic turbulence generated by the instability becomes smaller for $\sigma\gtrsim0.1$, which explains the observed decrease in acceleration efficiency, especially for ions (\fig{specsig15}(c) and (d), for $\sigma\gtrsim0.1$). 

The results presented in \fig{specsig15} refer to $\ompt=2250$. At later times, the nonthermal tail will extend to higher Lorentz factors in all cases, yet at a different rate for different magnetizations. Since ion acceleration is basically regulated by the degree of magnetic turbulence, which is the largest for $\sigma\sim0.1$, the fastest evolution in the ion spectral tail will be observed for $\sigma\sim0.1$ (black curve), which in fact extends to the highest energy. Instead, the electron tail evolves more rapidly for moderate  magnetizations ($\sigma=\pow{2}$ in blue), since electron acceleration in high-$\sigma$ flows is impeded by advection into the downstream (see \S\ref{sec:asub}). In general, the hierarchy in maximum energy observed  in \fig{specsig15} as a function of $\sigma$ corresponds to an analogous trend in the rates of acceleration.

Finally, we remark that the results reported here for $\theta=15\deg$ are  representative of the general behavior of low-obliquity shocks.

\begin{figure}[tbp]
\begin{center}
\includegraphics[width=0.5\textwidth]{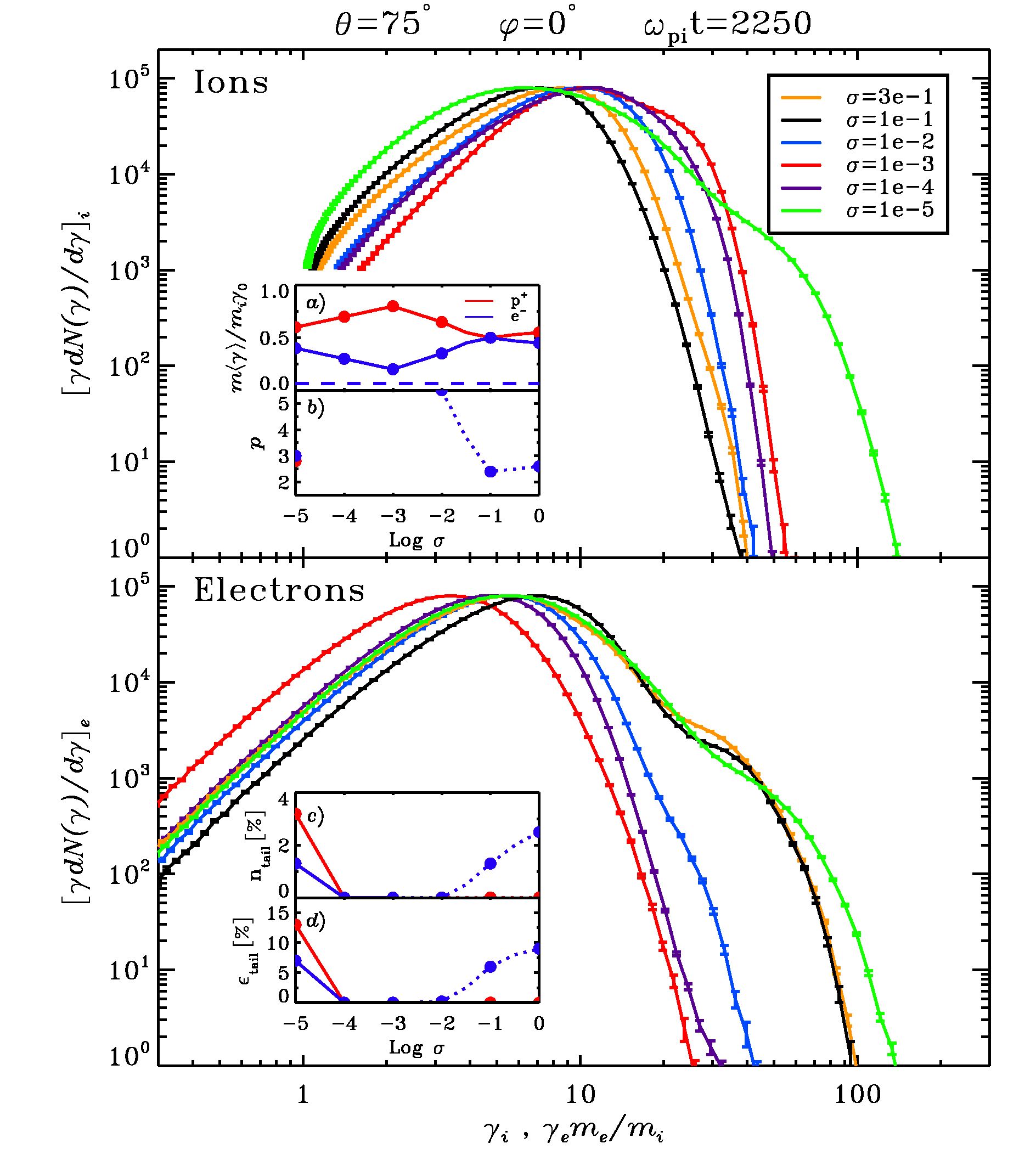}
\caption{Downstream particle spectrum at $\ompt=2250$ (upper panel for ions, lower panel for electrons) for different magnetizations. We fix the magnetic obliquity $\theta=75\deg$ (superluminal shock). Subpanels as in \fig{spectheta}.}
\label{fig:specsig75}
\end{center}
\end{figure}

%\vspace{-0.02in}
\subsubsection{Superluminal Shocks: $\thetacrit<\theta\leq90\deg$}\label{sec:msuper}
The downstream particle spectrum at different magnetizations for a superluminal shock with $\theta=75\deg$ is shown in \fig{specsig75}. The ion spectrum (upper panel) is always consistent with a purely Maxwellian distribution, with no evidence for nonthermal particles, except for the lowest magnetization ($\sigma=\pow{5}$, green curve). The non-Maxwellian shape of the ion spectrum for $\sigma=\pow{3}$ (red curve) results from incomplete ion thermalization at the shock front, and it will eventually relax to a Maxwellian distribution further downstream from the shock. In contrast, a prominent high-energy component appears in the electron spectrum (lower panel) for both very low (green for $\sigma=\pow{5}$) and very high (black for $\sigma=0.1$, yellow for $\sigma=0.3$) magnetizations, but not in between. However, the mechanism that regulates the electron high-energy tail is markedly different between low and high magnetizations, as we now describe.

In high-$\sigma$ flows ($\sigma\gtrsim0.1$; black curve for $\sigma=0.1$, yellow for $\sigma=0.3$), the electromagnetic precursor wave generated by the synchrotron maser instability ensures efficient transfer of energy from ions to electrons ahead of the shock, as discussed in \S\ref{sec:fsuper}. For $\sigma\gtrsim0.1$, the precursor is strong enough to drive electrons up to energy equipartition with ions (compare red and blue lines in panel (a) for $\sigma\gtrsim0.1$). As described in \S\ref{sec:asuper}, powerful wakefield oscillations excited by the precursor wave in the upstream region dissipate their energy by producing a pronounced high-energy tail in the electron spectrum (black for $\sigma=0.1$, yellow for $\sigma=0.3$). The tail contains $\sim2\%$ of electrons and $\sim8\%$ of electron energy (blue line in panels (c) and (d), respectively). Again, we remind that in superluminal magnetized shocks ($\sigma\gtrsim0.1$), the electron high-energy tail is due to wakefield \tit{heating} (rather than \tit{acceleration}), and it does \tit{not} extend in time to higher Lorentz factors (see \fig{spectime75}, for $\sigma=0.1$).

At intermediate magnetizations ($\pow{3}\lesssim\sigma\lesssim0.1$; blue curve for $\sigma=\pow{2}$, red for $\sigma=\pow{3}$), the synchrotron maser instability is gradually suppressed, since it is harder for the electron motion at the shock to stay coherent on the longer gyration time. It follows that the electromagnetic precursor is weaker \citep[][]{gallant_92}, and ion-to-electron energy transfer in the upstream is suppressed. As a result, the electron spectrum shifts to lower energies, whereas ions get hotter (see panel (a), from $\sigma=0.1$ down to $\sigma=\pow{3}$). A minor high-energy component, still due to wakefield heating, survives in the electron spectrum at $\sigma=\pow{2}$ (blue curve), but it disappears for $\sigma=\pow{3}$ (red curve). Despite the inefficiency of electron heating upstream of the shock, for $\sigma=\pow{3}$ the downstream electron population still contains as much as $\sim20\%$ of the initial ion kinetic energy (blue line in panel (a)). This is considerably in excess of $m_e/m_i\simeq6\%$ (dashed blue line in panel (a)), the value we would expect in the absence of any ion-to-electron energy transfer.\footnote{This result has been confirmed with larger mass ratios, up to $m_i/m_e=1000$.} In this regime, electrons gain energy not in the upstream region, but directly at the shock front by the cross-shock potential in the ion foot.

At lower magnetizations ($\pow{5}\lesssim\sigma\lesssim\pow{3}$; purple for $\sigma=\pow{4}$), the ion foot, whose scale is set by the Larmor radius of incoming ions in the shock-compressed field, becomes wider, so that electron heating by the cross-shock potential is more efficient (panel (a), for $\sigma\lesssim\pow{3}$). Finally, for $\sigma\lesssim\pow{5}$ (we have tested down to $\sigma=\pow{7}$), the ion foot is so wide that the counter-streaming between incoming and reflected ions can trigger the Weibel instability, before the upstream flow gets advected into the shock. As observed for unmagnetized shocks \citep[][]{spitkovsky_08}, this is accompanied by strong electron heating in the Weibel filaments and electron and ion acceleration via a Fermi-like mechanism. This explains the nonthermal tail seen in the electron and ion spectra of low-$\sigma$ flows (green curve for $\sigma=\pow{5}$). For ions, the tail is relatively flat ($p\sim2.7$, panel (b)) and it contains $\sim3\%$ of ions and $\sim13\%$ of ion energy (panels (c) and (d), respectively). The acceleration efficiency for electrons is smaller by roughly a factor of two (compare red and blue lines in panels (c) and (d), for $\sigma=\pow{5}$). Unlike in high-$\sigma$ shocks, here the ion and electron nonthermal tails extend with time to higher energies.

For $\sigma=\pow{5}$, the ion and electron spectra in \fig{specsig75} approach the results obtained for $\theta=15\deg$ at low magnetizations (e.g., $\sigma=\pow{3}$, red curve in \fig{specsig15}). Independently of magnetic obliquity, the shock now behaves as the unmagnetized case discussed by \citet{spitkovsky_08}. We find that the transition to the unmagnetized regime happens at lower magnetizations for higher obliquities ($\sigma\sim\pow{2}$ for $\theta=15\deg$, $\sigma\sim\pow{4}$ for $\theta=75\deg$), when the self-generated Weibel fields are comparable to the transverse component of the background field ($\propto\sqrt{\sigma}\sin\theta$). Neglecting the obliquity dependence, $\sigma\sim\pow{3}$ may be taken as a representative value for the transition point.

We remark that the results presented here for $\theta=75\deg$ are a good representation of the overall behavior of superluminal shocks, up to the perpendicular case $\theta=90\deg$.

%%%%%%%%%%%%%%%%%%%%%%%%%%%%%%%%%%%%
\begin{figure}[tbp]
\begin{center}
\includegraphics[width=0.5\textwidth]{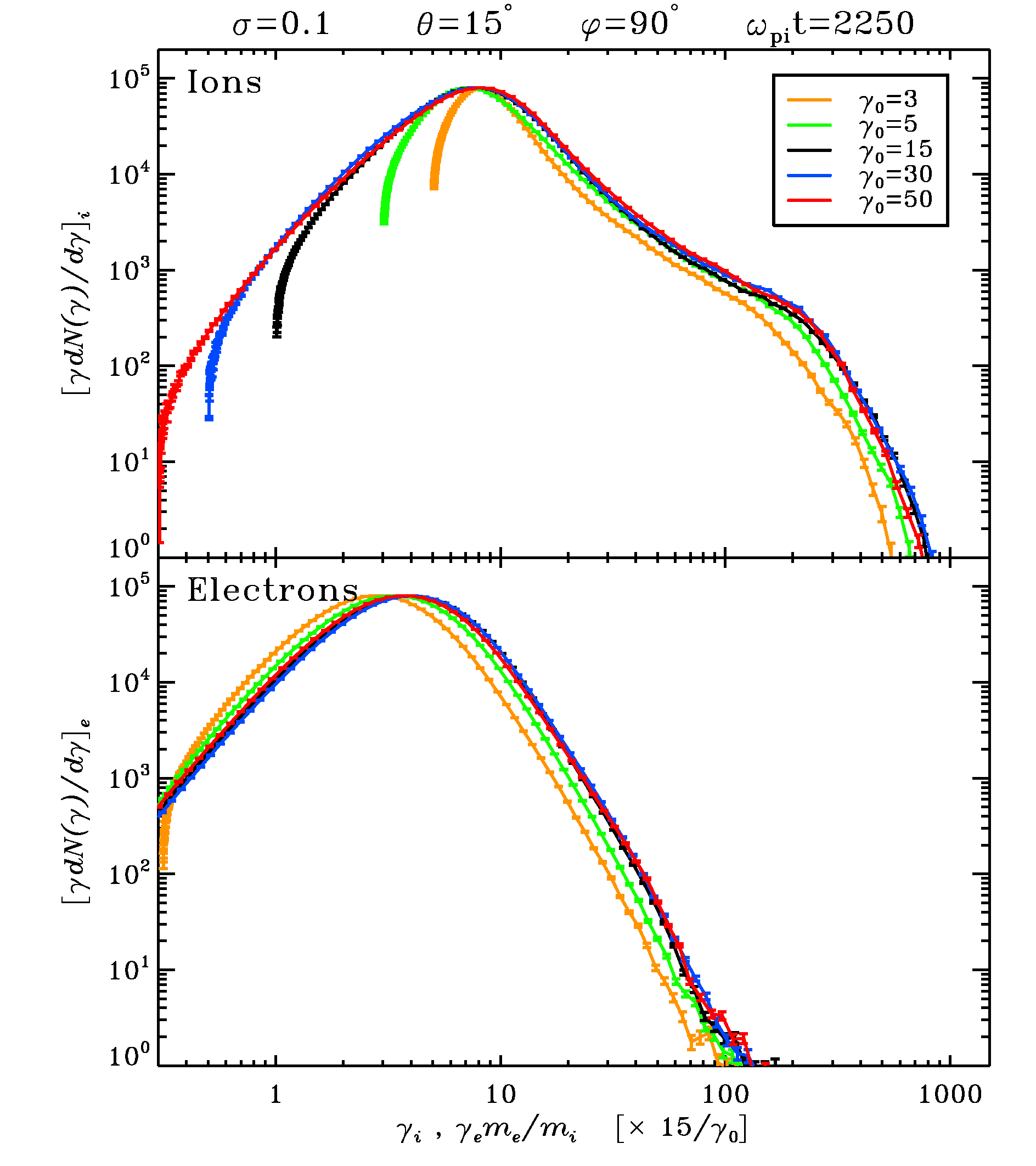}
\caption{Downstream particle spectrum at $\ompt=2250$ (upper panel for ions, lower panel for electrons) for different upstream bulk Lorentz factors. We fix the magnetic obliquity $\theta=15\deg$ (subluminal shock) and the magnetization $\sigma=0.1$. Spectra are shifted along the $x$-axis by $15/\gamma_0$ to facilitate comparison with the reference case $\gamma_0=15$.}
\label{fig:specgam15}
\end{center}
\end{figure}
\begin{figure}[tbp]
\begin{center}
\includegraphics[width=0.5\textwidth]{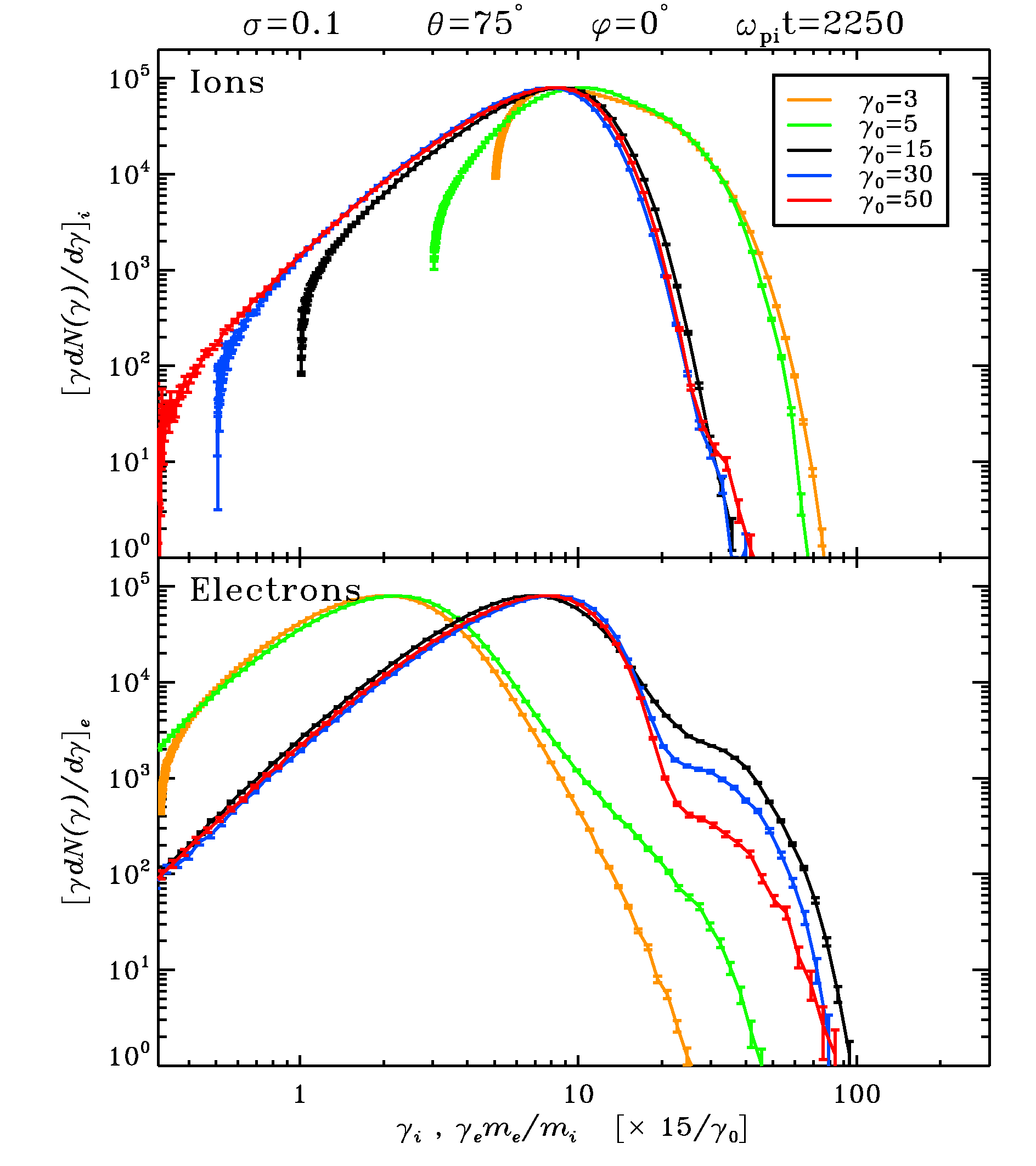}
\caption{Downstream particle spectrum at $\ompt=2250$ (upper panel for ions, lower panel for electrons) for different upstream bulk Lorentz factors. We fix the magnetic obliquity $\theta=75\deg$ (superluminal shock) and the magnetization $\sigma=0.1$. Spectra are shifted along the $x$-axis by $15/\gamma_0$ to facilitate comparison with the reference case $\gamma_0=15$.}
\label{fig:specgam75}
\end{center}
\end{figure}

%\vspace{0.25in}
\subsection{Dependence on Bulk Lorentz Factor $\gamma_0$}\label{sec:specgamma}
In this section, we keep the magnetization fixed at $\sigma=0.1$ and explore the dependence of our results upon the upstream bulk Lorentz factor, from $\gamma_0=3$ to $\gamma_0=50$. We analyze separately the case of subluminal ($\theta=15\deg$, in \S\ref{sec:gsub}) and superluminal ($\theta=75\deg$, in \S\ref{sec:gsuper}) shocks.

%\vspace{-0.03in}
\subsubsection{Subluminal Shocks: $0\deg \leq\theta<\thetacrit$}\label{sec:gsub}
\fig{specgam15} shows the downstream particle spectrum of a $\theta=15\deg$ subluminal shock, for different upstream bulk Lorentz factors. Spectra are shifted along the $x$-axis by $15/\gamma_0$, to facilitate comparison with the reference case $\gamma_0=15$ discussed in the previous sections. Once normalized in this way, the spectra for $\gamma_0\gtrsim15$ overlap almost perfectly (black for $\gamma_0=15$, blue for $\gamma_0=30$, red for $\gamma_0=50$), suggesting that the physics that regulates the acceleration of particles (both ions and electrons) is insensitive to the upstream bulk Lorentz factor, for highly relativistic flows. In fact, in our simulations we find that for $\gamma_0\gtrsim15$ the waves generated by Bell's instability have the same power and wavelength, once their power is normalized to the upstream bulk kinetic flux and the wavelength is measured in units of the upstream ion skin depth (or Larmor radius). The scaling seen in \fig{specgam15} for $\gamma_0\gtrsim15$ then follows from the fact that Bell's modes are of paramount importance for  particle acceleration in subluminal shocks (see \S\ref{sec:asub_mech}).

Spectra for lower Lorentz factors (green for $\gamma_0=5$, yellow for $\gamma_0=3$) still share the same overall properties of $\gamma_0=15$, albeit with a tendency for lower high-energy cutoffs and cooler electron distributions. Overall, the properties of $\gamma_0\gtrsim3$ subluminal shocks seem to be well captured by our representative case $\gamma_0=15$. 

In mildly relativistic magnetized flows ($\gamma_0\lesssim2$, for $\sigma=0.1$), the shock is mediated by a different process, such that the dominant mode is now generated by gyro-frequency resonance with the returning ions, rather than by Bell's non-resonant instability  (see Appendix \ref{sec:comparison}). This may change the efficiency of particle acceleration. Such trans-relativistic regime is not the subject of this work and will be addressed elsewhere.

\subsubsection{Superluminal Shocks: $\thetacrit<\theta\leq90\deg$}\label{sec:gsuper}
As shown in \fig{specgam75}, the spectrum of $\theta=75\deg$ superluminal shocks presents a much more complicated dependence on the upstream bulk Lorentz factor. The ion spectrum (upper panel) is consistent with a thermal distribution in the whole range of $\gamma_0$ we explore (from $\gamma_0=3$ to $\gamma_0=50$), but for $\gamma_0\lesssim5$ (green for $\gamma_0=5$, yellow for $\gamma_0=3$) the ion Maxwellian is much hotter than for $\gamma_0\gtrsim10$ (black for $\gamma_0=15$, blue for $\gamma_0=30$, red for $\gamma_0=50$). Correspondingly, the electron distribution (lower panel), which peaks at roughly the same energy as ions for $\gamma_0\gtrsim10$, shifts to much lower energies for $\gamma_0\lesssim5$.

\citet{lyubarsky_06} argues that  the synchrotron maser instability should only operate for $\gamma_0\gtrsim(m_i/m_e)^{1/3}$. In this regime, the shock-compressed field grows on a scale smaller than the electron Larmor radius, and the non-adiabatic motion of incoming electrons into the shock will result in a ring-like distribution in phase space, which is synchrotron-maser unstable. It follows that a powerful electromagnetic precursor will only be generated in high-$\gamma_0$ shocks. Our results show that, for $m_i/m_e=16$, the transition occurs at $\gamma_0\sim10$. For $\gamma_0\gtrsim10$, when the precursor wave is strong, electrons reach energy equipartition with ions ahead of the shock. As discussed in \S\ref{sec:asuper}, powerful electrostatic oscillations generated in the wake of the precursor dissipate their energy by heating a fraction of the incoming electrons, that eventually populate the high-energy component seen for $\gamma_0\gtrsim10$ in the electron spectrum of \fig{specgam75}. Instead, for $\gamma_0\lesssim10$, wakefield oscillations are much weaker, so that the electron high-energy component tends to disappear (green for $\gamma_0=5$, yellow for $\gamma_0=3$). Also, since the precursor is less powerful in low-$\gamma_0$ flows, ion-to-electron energy transfer in the upstream is suppressed, and the downstream electron spectrum shifts to lower energies. Yet, the downstream electron population still contains an appreciable fraction ($\sim15\%$) of the initial ion energy, due to significant electron energization by the cross-shock potential. 

A note of caution is required concerning the electron high-energy tail for $\gamma_0\gtrsim10$. As shown in \fig{spectime75}, this component grows in time up to a maximum (reached between $\ompt\sim2000$ and $\ompt\sim3000$), and then it recedes (although we cannot exclude a cyclical behavior). The growth time scales with the wavelength of wakefield oscillations, which is longer (in units of $\comp$) for larger bulk Lorentz factors \citep{lyubarsky_06}. It follows that at fixed $\ompt=2250$, while $\gamma_0=15$ (black curve in \fig{specgam75}) is already at maximum, the electron high-energy population is still building up for $\gamma_0=30$ (blue curve) and $\gamma_0=50$ (red curve). At later times (not shown in \fig{specgam75}), the electron spectrum for $\gamma_0=30$ and $\gamma_0=50$ will roughly resemble the case $\gamma_0=15$ presented here. 

In summary, our results for  $\gamma_0=15$ are a good proxy for highly relativistic flows ($\gamma_0\gtrsim10$), but they do not apply to mildly relativistic shocks.

%%%%%%%%%%%%%%%%%%%%%%%%%%%%%%%%%%%%
\section{Summary and Discussion}\label{sec:disc}
We have explored by means of 2.5D PIC simulations the internal structure and acceleration properties of relativistic collisionless shocks propagating in a magnetized electron-ion plasma. This work complements the study of SS09, that analyzed relativistic shocks in magnetized electron-positron plasmas. To investigate how the shock properties depend on the conditions  of the upstream flow, we vary the bulk Lorentz factor $\gamma_0$, the magnetization $\sigma$ and the angle $\theta$ between the upstream magnetic field and the shock normal. Our results, confirmed by extensive convergence studies, can be summarized as follows:

\begin{list}{\labelitemi}{\leftmargin=1em}
\item For $\sigma\sim0.1$, the magnetic field is so strong that charged particles are constrained to move along the field lines, which are advected downstream from the shock. If the magnetic obliquity $\theta<\thetacrit\simeq34\deg$ (``subluminal'' shocks), downstream particles can return upstream by following the magnetic field. In this case, the electric current of ``returning'' ions can trigger the so-called Bell's instability \citep{bell_04}, which generates powerful circularly-polarized Alfv\'enic-type waves in the upstream. The shock results from the nonlinear steepening of such waves, which happens quasi-periodically every few ion gyro-times (the so-called ``shock reformation'' process). 
Instead, for $\theta>\thetacrit$ (``superluminal'' shocks), no particle can propagate ahead of the shock along the field, and counter-streaming instabilities are suppressed. In this case, the shock is mediated by magnetic reflection of the incoming flow off the shock-compressed field \citep[e.g.,][]{alsop_arons_88}. In short, the dichotomy between subluminal and superluminal configurations, presented by SS09 for electron-positron flows, still holds in electron-ion shocks.

\item If particle acceleration requires repeated bounces back and forth across the shock, as in the standard Fermi picture, only \tit{subluminal} shocks should be efficient accelerators. Or, as measured in the upstream frame, the magnetic field needs to be within an angle $\sim\thetacrit/\gamma_0\simeq34\deg/\gamma_0$ from the shock normal. In subluminal shocks, particles are accelerated mostly via non-resonant interactions with Bell's waves (a Fermi-like diffusive process), although for $30\deg\lesssim\theta\lesssim\thetacrit$ acceleration by the upstream motional electric field may be important as well (Shock Drift/Surfing Acceleration; see SS09 for similar conclusions). In the downstream spectrum, the shock-accelerated particles populate a power-law nonthermal tail. For our representative case $\theta=15\deg$, the ion tail contains $\sim5\%$ of ions and $\sim30\%$ of ion energy, and its slope is $-2.1\pm0.1$. The tail stretches linearly with time to higher Lorentz factors. Electrons enter the shock with a significant fraction ($\sim15\%$) of the initial ion energy, yet their Larmor radius is still small compared to ions, and they are rapidly advected downstream with the magnetic field. It follows that the acceleration of electrons is less efficient ($\sim2\%$ by number and $\sim10\%$ by energy, with a steep tail of slope $-3.5\pm0.1$).  In the upstream, the pressure of shock-accelerated ions (in fact, few electrons propagate back upstream) can significantly alter the structure of the shock, forming a smooth density ``precursor'' ahead of the actual ``subshock'' (this is the characteristic profile of ``cosmic-ray modified'' shocks). The overall width of the shock layer increases with time at $\simeq0.1\,c$.

\item In \tit{superluminal} shocks, the synchrotron maser instability \citep{hoshino_91} generates a train of transverse electromagnetic ``precusor'' waves propagating into the upstream, that push on the incoming electrons, making them lag behind ions. This initiates longitudinal ``wakefield'' oscillations in the upstream, that transfer energy from ions to electrons \citep{lyubarsky_06}. In high-obliquity shocks, electrons enter the shock with bulk energy comparable to ions. So, electrons acquire most of their energy  in the \tit{upstream} rather than at the \tit{shock}, in contrast to what is assumed in models that only consider the cross-shock potential  \citep[e.g.,][]{gedalin_08}. A few percent of the incoming electrons receive an extra amount of heating from the dissipation of wakefield oscillations. They eventually populate a prominent high-energy feature in the downstream electron spectrum, which may resemble a (short) power-law tail. However, we remark that it results from \tit{heating}, and not \tit{acceleration} \citep[as argued by][]{hoshino_08}, and it does \tit{not} extend in time to higher energies. Rather, as the incoming electrons get hotter, the generation of electromagnetic precursor waves is suppressed \citep{hoshino_91}, and the electron high-energy component recedes to lower energies. Hereafter, a cyclical behavior is likely, though undemonstrated. Despite the significant wakefield heating experienced in the upstream, which may facilitate injection into a diffusive acceleration process, no signature of Fermi-accelerated electrons is seen in magnetized superluminal shocks (see also SS09).

\item The properties of magnetized subluminal shocks do not significantly depend on the upstream bulk Lorentz factor, in the relativistic regime $\gamma_0\gtrsim3$ we explore. For superluminal angles, the synchrotron maser instability is suppressed for $\gamma_0\lesssim10$ \citep[][]{lyubarsky_06}. So, for low-$\gamma_0$ shocks the electromagnetic precursor is weaker, and transfer of energy from ions to electrons in the upstream is much less efficient. Yet, downstream electrons still account for as much as $\sim15\%$ of the initial ion energy, due to significant electron energization by the cross-shock potential in the ion foot.

\item For lower magnetizations ($\sigma\lesssim10^{-3}$), the incoming plasma can filament, and the shock is mediated by the ion Weibel instability \citep{weibel_59, medvedev_loeb_99,gruzinov_waxman_99}, which ensures strong electron heating and efficient particle acceleration, for both ions and electrons \citep{spitkovsky_08,martins_09}. The nonthermal tail in the downstream ion spectrum contains $\sim5\%$ of ions and $\sim20\%$ of ion energy, and its slope is $-2.5\pm0.1$. A similar tail is present in the electron spectrum, with comparable slope but a smaller particle ($\sim2\%$) and energy ($\sim10\%$)  content.  For magnetizations $\sigma\lesssim10^{-3}$, the fields generated by the Weibel instability are stronger than the background field, and no difference persists between subluminal and superluminal configurations.

\item An important question is whether our 2.5D simulations can capture the relevant three-dimensional physics of electron-ion shocks. We performed a limited number of 3D experiments with relatively small computational grids ($\sim3$ ion skin depths along each transverse dimension), obtaining essentially the same results as in our 2.5D simulations. Yet, large 3D runs are certainly desirable to confirm  the overall picture presented here.
\end{list}

Our findings may constrain the composition, Lorentz factor and magnetization of jets in Active Galactic Nuclei (AGNs) and Gamma-Ray Bursts (GRBs).\footnote{Pulsar Winds are thought to be dominated by electron-positron pairs, and they will not be discussed here (we refer to SS09). If a significant ionic component is present, electron (and positron) acceleration may proceed via the Resonant Cyclotron Absorption mechanism discussed by \citet{hoshino_92} and \citet{amato_arons_06}.} 
Synchrotron and inverse Compton emission from the core of blazar jets is usually attributed to high-energy electrons accelerated in mildly-relativistic internal shocks ($\gamma_0\sim2$ in the jet comoving frame). Such shocks should be quasi-perpendicular, since polarization measurements of radio knots \citep{gabuzda_04, pushkarev_05} indicate that the magnetic field is mostly transverse to the jet axis. Electron acceleration in internal shocks is also invoked to explain the prompt emission of GRBs \citep[e.g.,][]{piran_04}. If the jet flow carries a substantial toroidal field, internal shocks in GRBs will be magnetized and quasi-perpendicular. For both AGNs and GRBs, it is inferred that the emitting electrons should contain a substantial fraction ($\epsilon_e\sim10\%$) of the flow energy, and their spectrum is usually modeled as a power law extending over several decades in energy.

We find that significant heating of electrons is a universal property of relativistic magnetized shocks. The mechanism involved may vary depending on magnetization and magnetic obliquity, but post-shock electrons account for at least $\sim20\%$ of the bulk pre-shock energy, regardless of the upstream conditions. Although the agreement between this lower limit and the observational requirement mentioned above is surely encouraging, we remark that most of our downstream electrons belong to the \tit{thermal} component, whereas the emitting particles from AGNs and GRBs are usually thought to be \tit{nonthermal}. Yet, our findings are important to constrain the characteristic downstream electron energy to be $\gtrsim 0.2\,\gamma_0 m_ic^2$, with the following two implications. First, this rules out models of GRB emission \citep[e.g.,][]{bykov_96} which assume that only a small fraction of electrons are accelerated up to high energies (and share the whole $\epsilon_e$), whereas the bulk of electrons remain cold (at $\sim\gamma_0m_ec^2$). Second, the minimum Lorentz factor of a hypothetical power-law tail will be $\gtrsim 0.2\,\gamma_0 m_i/m_e$. For models of GRB and AGN emission that require a power-law distribution starting from smaller Lorentz factors, this would suggest that electron-positron pairs may be a major component of the flow (see SS09).

Regarding nonthermal acceleration of electrons, we find that quasi-perpendicular magnetized shocks are inefficient particle accelerators, in contrast to what is required in models of GRBs and AGN jets. However, the high-energy component that we find in the electron spectrum of high-obliquity shocks, which actually results from heating (and not acceleration), may be (mis)interpreted as a short power-law tail. In fact, this component might reasonably explain the emission from AGN jets, which often requires an electron distribution extending only for one decade in energy \citep{celotti_08}. Otherwise, our findings imply that the jet pre-shock flow should be weakly magnetized ($\sigma\lesssim\pow{3}$), or maybe seeded with strong small-scale magnetic turbulence, that could facilitate the Fermi process even in superluminal shocks \citep[][]{sironi_goodman_07}. Alternatively, electron acceleration may occur not in shocks but in reconnection layers of a Poynting-dominated jet  \citep[e.g.,][]{lyutikov_03,giannios_06}.

Finally, if internal shocks in AGNs and GRBs are the sources of High Energy Cosmic Ray ions \citep[e.g.,][]{sironi_socrates_10}, then the pre-shock magnetic field needs to be nearly aligned with the shock normal, within an angle $\sim\thetacrit/\gamma_0\simeq34\deg/\gamma_0$ (as measured in the upstream fluid frame), for the shock to be ``subluminal''. Alternatively, the acceleration of cosmic ray protons may occur at weakly magnetized shocks, like GRB external shocks or the termination shock (``hot spot'') of AGN jets, or in the jet ``sheath'', where interaction of the jet with the surrounding medium may create subluminal configurations. In any case, we find that ion acceleration is always accompanied by substantial magnetic field amplification, especially ahead of the shock. In the presence of accelerated electrons, the upstream region may then contribute considerably to the observed emission. Synthetic spectra extracted from PIC simulations of shocks \citep[as done in unmagnetized pair shocks by][]{sironi_spitkovsky_09b} may possibly reveal characteristic signatures that could observationally distinguish such \tit{upstream} emission from the \tit{downstream}-based models that are usually employed.

\vspace{0.1in}
\tit{Acknowledgements.} We thank J.~Arons, A.~Beloborodov, R.~Blandford, L.~Gargate, D.~Giannios, M.~Lyutikov, M.~Medvedev, G.~Pelletier, M.~Riquelme and R.~Romani for insightful comments. LS gratefully thanks the Osservatorio Astronomico di Brera for hospitality and G.~Ghisellini for stimulating discussions. This research was supported by NSF grant AST-0807381 and NASA grant NNX10AI19G. The simulations presented in this article were performed on computational resources supported by the PICSciE-OIT High Performance Computing Center and Visualization Laboratory. This research used resources of the National Energy Research Scientific Computing Center, which is supported by the Office of Science of the U.S. Department of Energy under Contract No. DE-AC02-05CH11231.

%\newpage
%\vspace{0.5in}
\appendix

%%%%%%%%%%%%%%%%%%%%%%%%%%%%%%%%%%%%%%%%
\section{A) Dependence on the Mass Ratio}\label{sec:specmime}
\begin{figure}[tbp]
\begin{center}
\includegraphics[width=0.5\textwidth]{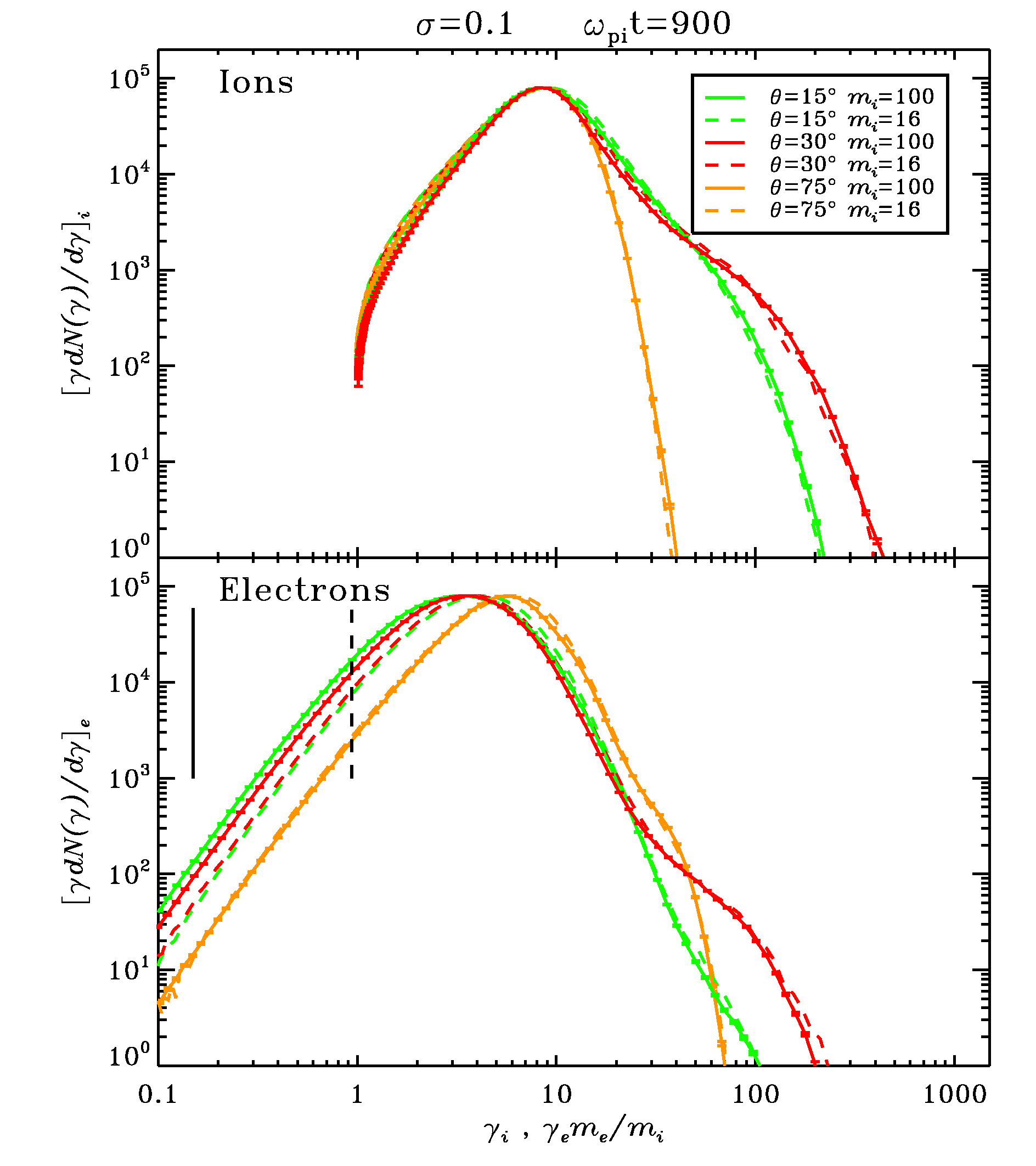}
\caption{Comparison of downstream particle spectra (upper panel for ions, lower panel for electrons) between different mass ratios: $m_i/m_e=16$ (dashed lines) and $m_i/m_e=100$ (solid lines). The black vertical lines (dashed for $m_i/m_e=16$, solid for $m_i/m_e=100$) correspond to $\gamma_0 m_e/m_i$, i.e., the location expected for the peak of the electron distribution in the absence of any ion-to-electron energy transfer.}
\label{fig:specmime}
\end{center}
\end{figure}
The reliability of our results may be questioned based on the artificially small mass ratio ($m_i/m_e=16$) that we employ in most of our simulations. However, as we show here, for the parameters considered in the present work, a mass ratio of $m_i/m_e=16$ is already large enough to separate the electron and ion scales, and capture the relevant physics of electron-ion shocks.

\fig{specmime} compares the downstream particle spectra for three representative magnetic obliquities ($\theta=15\deg$, $30\deg$ and $75\deg$, with $\gamma_0=15$ and $\sigma=0.1$) between two values of the mass ratio: $m_i/m_e=16$ (dashed lines) and $m_i/m_e=100$ (solid lines). The spectra are computed at the same time (in units of $\omega_{\rm pi}^{-1}$) and at the same distance behind the shock (in units of $\comp$). Very good agreement is obtained in all cases, for both ions and electrons. 

The downstream electron spectrum peaks at much higher energies than expected in the absence of any ion-to-electron energy transfer (compare with black vertical lines; dashed for $m_i/m_e=16$, solid for $m_i/m_e=100$), and the location of the peak is roughly independent of the mass ratio.  It follows that the ``effective'' electron mass, including the average electron Lorentz factor, is a substantial fraction of $\gamma_0m_i$, independent of $m_i/m_e$ (for $m_i/m_e\gtrsim16$). The electron physics will then evolve on the ion time and length scales, which explains the similarity in \fig{specmime} between  electron spectra with different $m_i/m_e$, in terms of both the low-energy thermal part and the high-energy component. 

Various tests, for mass ratios up to $m_i/m_e=1000$ and for different upstream parameters (we changed both $\sigma$ and $\gamma_0$), confirm the overall picture presented here.

%%%%%%%%%%%%%%%%%%%%%%%%%%%%%%%%%%%%%%%%
\section{B) Dependence on the Field Orientation with respect to the Simulation Plane}\label{sec:specphi}
\begin{figure}
\centering
\subfigure[]{
\includegraphics[width=0.47\textwidth]{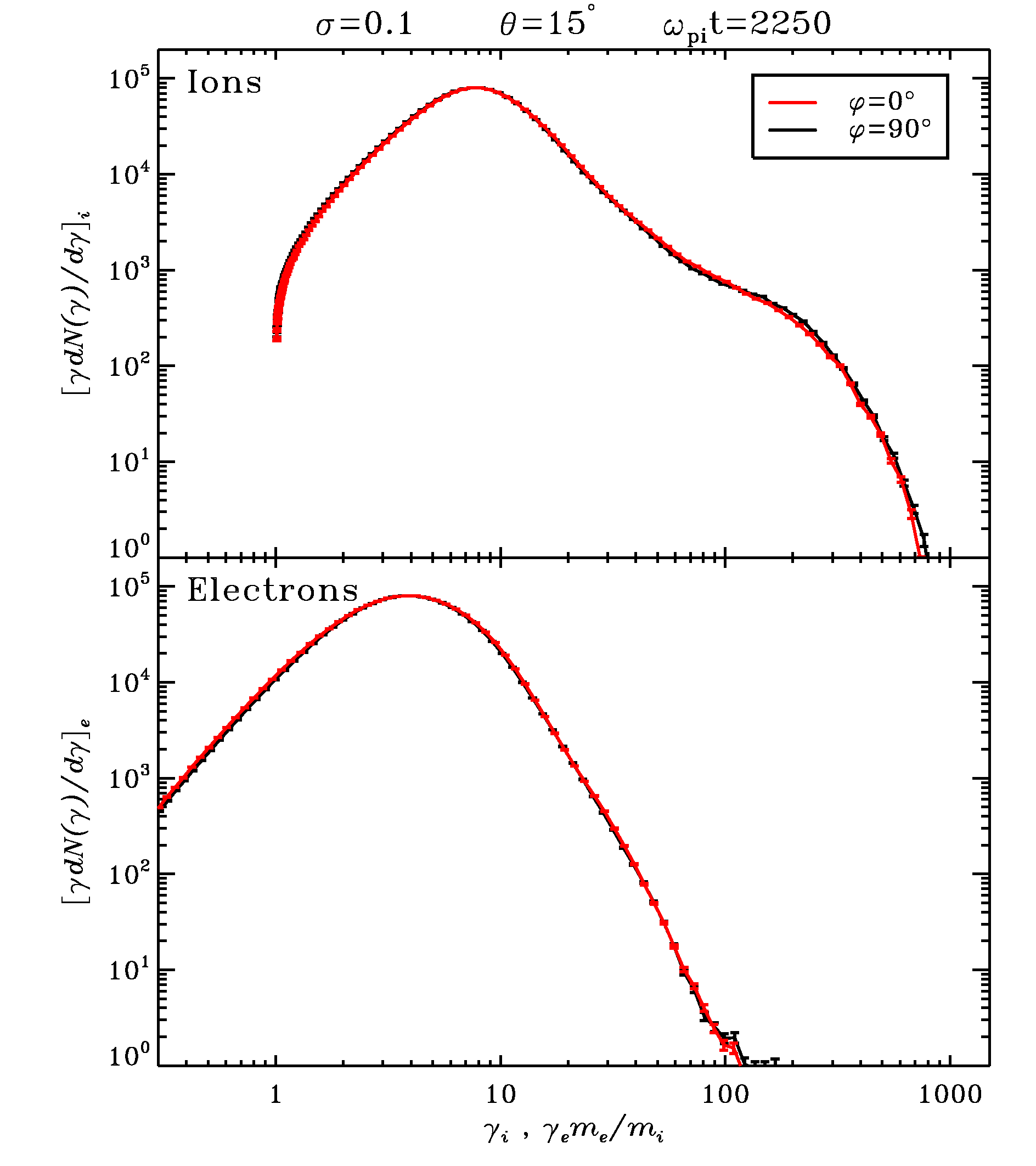}
}
\hspace{0.1in}
\subfigure[]{
\includegraphics[width=0.47\textwidth]{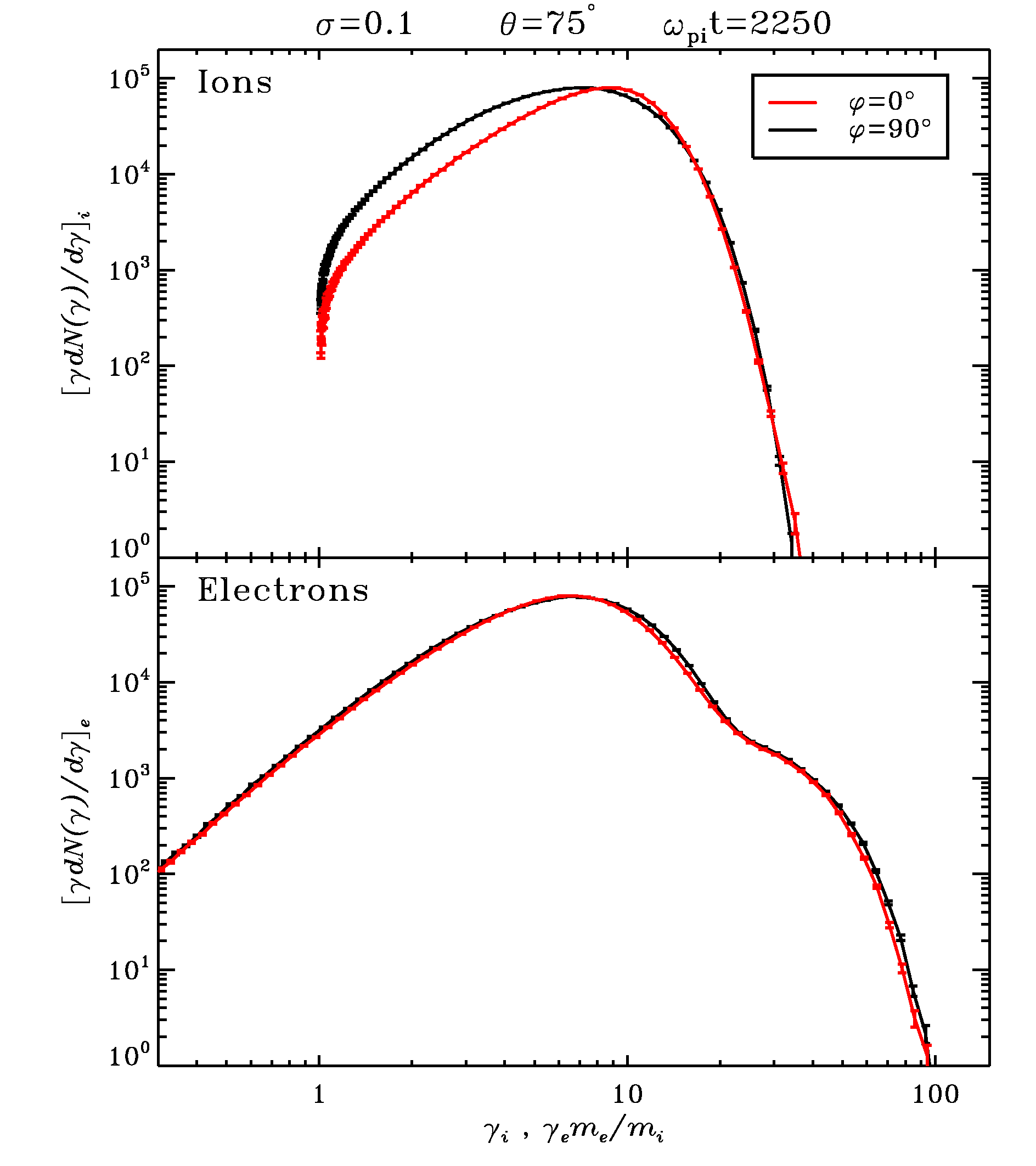}
}
\caption{For fixed magnetic obliquity ($\theta=15^\circ$ in panel (a), $\theta=75^\circ$ in panel (b)), downstream particle spectra at $\ompt=2250$ for different orientations of the upstream magnetic field $\mathbf{B}_0$ with respect to the simulation plane: magnetic field lying either in the simulation plane ($\varphi=0\deg$, in red) or in a plane perpendicular to the simulation plane ($\varphi=90\deg$, in black).}
\label{fig:specphi}
\end{figure}
We now assess the robustness of our results upon the orientation of the upstream magnetic field $\mathbf{B}_0$ with respect to the 2D simulation plane. In \fig{specphi} (upper panel for ions, lower panel for electrons), we plot with red lines the downstream spectrum  if $\mathbf{B}_0$ is in the simulation plane (i.e., $\varphi=0\deg$), whereas black lines are for $\mathbf{B}_0$ lying in a plane perpendicular to the simulation plane  (i.e., $\varphi=90\deg$). 

For low-obliquity shocks (e.g., $\theta=15\deg$ in panel (a)), our results are independent on the angle $\varphi$. However, for shocks in which the obliquity is close to the critical boundary $\thetacrit\simeq34\deg$ between subluminal and superluminal configurations, we observe a substantial suppression of SDA for in-plane magnetic fields, both in electron-positron and electron-ion flows (see $\theta=30\deg$ in Fig.~21(b) of SS09, for pair shocks). As discussed in SS09, the case with $\varphi=90\deg$ is in better agreement with 3D simulations. For this reason, in the whole range of subluminal angles presented in the current work, we employ $\varphi=90\deg$.

For high-obliquity shocks (e.g., $\theta=75\deg$ in panel (b)), the agreement between in-plane and out-of-plane results is also remarkably good. The only appreciable difference is in the low-energy part of the ion spectrum (upper right panel). For $\varphi=90\deg$, the plane perpendicular to the magnetic field (where downstream particle orbits would tend to lie) is almost degenerate with the simulation plane, resulting in a downstream ion distribution which is closer to a 2D (rather than 3D) Maxwellian. Instead, for $\varphi=0\deg$ the two planes are well separated, and ion isotropization in three dimensions is favored. For this reason, we preferentially employ $\varphi=0\deg$ for superluminal shocks.

%%%%%%%%%%%%%%%%%%%%%%%%%%%%%%%%%%%%%%%%
\section{C) Ion Counter-Streaming Instabilities}\label{sec:comparison}
In both unmagnetized and subluminal magnetized shocks, the interaction between the incoming flow and the shock-accelerated particles that propagate back upstream is a potential source of counter-streaming plasma instabilities. In this section, we determine how the nature of the instability depends on the upstream parameters (bulk Lorentz factor $\gamma_0$ and magnetization $\sigma$, choosing $\theta=0\deg$ as a representative subluminal obliquity). Primed quantities are measured in the upstream fluid frame, unprimed quantities in the downstream (simulation) frame.

Let $\zeta_{\msc{cr}}\equiv n_{\msc{cr}}/n_i$ be the ratio of cosmic-ray number density to number density of incoming ions. By ``cosmic rays'', we mean the shock-accelerated ions moving ahead of the shock (we neglect the returning electrons, since their contribution is usually smaller). In the upstream frame, $\zeta'\crs=\gamma_0^2\,\zeta\crs$, given that the population of returning ions is roughly isotropic in the simulation frame. Similarly, we define $\epsilon\crs\equiv U\crs/(\gamma_0m_i n_i c^2)$ as the ratio of cosmic ray energy density to kinetic energy density of the injected ions, such that the average cosmic ray Lorentz factor is $\gamma\crs=\gamma_0(\epsilon\crs/\zeta\crs)$. In the upstream frame, $\epsilon'\crs=\gamma_0^4\epsilon\crs$ and $\gamma'\crs=\epsilon'\crs/\zeta'\crs=\gamma_0\,\gamma\crs$. Informed by Figs.~\ref{fig:specsig15} and \ref{fig:specgam15}, we can assume to a first approximation that $\zeta\crs$ and $\epsilon\crs$ (as measured in the downstream frame) do not significantly depend on $\gamma_0$ or $\sigma$.

The maximum growth rate of Bell's instability ($\omega'_{\rm Bell}\simeq1/2\,\zeta'\crs\,\omega_{\rm pi}$) is achieved for a characteristic wavelength $\lambda'_{\rm Bell}\simeq4\pi(\sqrt{\sigma'}/\zeta\crs)\,c/\omega_{\rm pi}$  \citep{reville_06}. Here, $\sigma'$ is the magnetization parameter as defined in the upstream frame ($\sigma'=\gamma_0^2\,\sigma$, for $\theta=0\deg$). The corresponding Larmor frequency of background ions is $\omega'_{\rm{ci}}=\gamma_0\,\omega_{\rm{ci}}$.

Bell's instability governs the evolution of the shock if the following two conditions are simultaneously satisfied: (\tit{i}) its characteristic wavelength $\lambda'_{\rm Bell}$ is smaller than $r'_{\msc{l,cr}}$, the Larmor radius of cosmic ray ions; (\tit{ii}) its growth rate is smaller than the Larmor frequency of background ions ($\omega'_{\rm Bell}\lesssim\omega'_{\rm{ci}}$), i.e., the upstream plasma is magnetized. Since $r'_{\msc{l,cr}}=\gamma'\crs/\sqrt{\sigma'}\,\comp$, the first condition can be rewritten as $2\,\sigma'\lesssim\epsilon'\crs$ in the upstream fluid frame, which corresponds, in terms of quantities in the simulation frame, to
\begin{equation}\label{eq:c1}
2\,\sigma\lesssim\gamma_0^2\,\epsilon\crs~~.
\end{equation}
When the above inequality is broken (either for large $\sigma$ or small $\gamma_0$, at fixed $\epsilon\crs$), the physics of the instability changes. The dominant mode is now generated via gyro-frequency resonance with the cosmic rays (as opposed to Bell's instability, which produces non-resonant modes). We observe this change of polarization in the dominant mode for magnetizations $\sigma\gtrsim0.3$ (with fixed $\gamma_0=15$) and for bulk Lorentz factors $\gamma_0\lesssim2$ (with fixed $\sigma=0.1$).

The condition (\tit{ii}), that the upstream ions stay magnetized on the growth time of Bell's instability, can be rewritten as $\zeta'\crs\lesssim\sqrt{\sigma'}$ in the upstream frame. In the simulation frame, 
\begin{equation}\label{eq:c2}
\gamma_0\,\zeta\crs\lesssim\sqrt{\sigma}~~.
\end{equation}
For lower magnetizations, or higher bulk Lorentz factors, the upstream plasma can filament, and the ion Weibel instability dominates. For $\gamma_0=15$, this transition happens in our simulations around $\sigma\sim10^{-2}$. 

\bibliography{ions}
\end{document}